\documentclass[%
 reprint,
 superscriptaddress,preprintnumbers,
 nofootinbib,
 amssymb,
 aps,
]{revtex4-2}

\usepackage[utf8]{inputenc}
\usepackage[colorlinks=true,citecolor=blue,linkcolor=blue]{hyperref}
\usepackage{amsmath,amssymb,mathtools}
\usepackage{graphicx}               
\usepackage{url}
\usepackage[dvipsnames]{xcolor}
\usepackage{xspace}
\usepackage{dsfont}

\usepackage[noindentafter]{titlesec}
\allowdisplaybreaks
\DeclareMathVersion{sans}
\SetSymbolFont{operators}{sans}{OT1}{cmbr}{m}{n}
\SetSymbolFont{letters}{sans}{OML}{cmbrm}{m}{it}
\SetSymbolFont{symbols}{sans}{OMS}{cmbrs}{m}{n}
\SetMathAlphabet{\mathit}{sans}{OT1}{cmbr}{m}{sl}
\SetMathAlphabet{\mathbf}{sans}{OT1}{cmbr}{bx}{n}
\SetMathAlphabet{\mathtt}{sans}{OT1}{cmtl}{m}{n}
\SetSymbolFont{largesymbols}{sans}{OMX}{iwona}{m}{n}

\DeclareMathVersion{boldsans}
\SetSymbolFont{operators}{boldsans}{OT1}{cmbr}{b}{n}
\SetSymbolFont{letters}{boldsans}{OML}{cmbrm}{b}{it}
\SetSymbolFont{symbols}{boldsans}{OMS}{cmbrs}{b}{n}
\SetMathAlphabet{\mathit}{boldsans}{OT1}{cmbr}{b}{sl}
\SetMathAlphabet{\mathbf}{boldsans}{OT1}{cmbr}{bx}{n}
\SetMathAlphabet{\mathtt}{boldsans}{OT1}{cmtl}{b}{n}
\SetSymbolFont{largesymbols}{boldsans}{OMX}{iwona}{bx}{n}

\def\thesection{\arabic{section}}
\def\thesubsection{\arabic{section}.\arabic{subsection}}
\def\thesubsubsection{\arabic{section}.\arabic{subsection}.\arabic{subsubsection}}
\titleformat{\section}{\centering \large \mathversion{boldsans} \sffamily\bfseries \color{blue!80!black} }{\thesection}{15pt}{}{}
\titlespacing{\section}{0pt}{15pt}{5pt}
\titleformat{\subsection}{\normalsize \sffamily \mathversion{sans} \color{blue!70!black} }{\thesubsection}{10pt}{}{}
\titlespacing{\subsection}{0pt}{10pt}{5pt}
\titleformat{\subsubsection}{\normalsize \sffamily \mathversion{sans} \itshape  \color{blue!70!black} }{\thesubsubsection}{10pt}{}{}
\titlespacing{\subsubsection}{0pt}{10pt}{3pt}

\makeatletter
    \renewcommand{\p@subsection}{}
    \renewcommand{\p@subsubsection}{}
\makeatother

\makeatletter
\let\MyIntOrig\int
\def\MyIntSpace{\hspace{-.35em}} 
\def\int{\MyInt}
\def\MyInt{\MyIntOrig\MyIntSkipMaybe}
\def\MyIntSkipMaybe{
	\@ifnextchar_{\MyIntSkipScript}{%
		\@ifnextchar^{\MyIntSkipScript}{%
			\@ifnextchar\limits{\MyIntSkipTok}{%
				\@ifnextchar\nolimits{\MyIntSkipTok}{%
					\MyIntSpace}}}}%
}
\def\MyIntSkipScript#1#2{#1{#2}\MyIntSkipMaybe}
\def\MyIntSkipTok#1{#1\MyIntSkipMaybe}

\definecolor{red}{rgb}{0.6,.0706,.1373}
\definecolor{blue}{rgb}{0,0.396,0.741}
\definecolor{orange}{rgb}{.8, .4806, 0.173}

\colorlet{blueref}{blue!80!black}
\colorlet{bluelink}{blue!90!black}

\hypersetup{
    citecolor=blueref, 	  
    linkcolor=bluelink,	  
    urlcolor=bluelink,	  
    colorlinks = true
}

\newcommand{\eminus}{\vcenter{\hbox{\scalebox{0.6}[1]{$ - $}}}}	
\newcommand{\ord}[1]{\mathcal{O}( #1 )}
\newcommand{\sscript}[1]{{\scriptscriptstyle \mathrm{#1}}}

\newcommand{\dd}{\mathop{}\!\mathrm{d}}
\newcommand{\transpose}{^\intercal}

\newcommand{\UV}{\sscript{UV}}
\newcommand{\EFT}{\sscript{EFT}}

\newcommand{\soft}[1]{{\color{red!80!white} #1}}
\newcommand{\hard}[1]{{\color{violet!70!blue!90!white} #1}}
\newcommand{\msbar}{$ \overline{\text{\small MS}} $\xspace}

\begin{document}

\title{\texorpdfstring{\sffamily \color{blue!80!black} \LARGE}{} Functional Matching and Renormalization Group Equations\texorpdfstring{\\}{} at Two-Loop Order}

\author{Javier Fuentes-Mart\'{\i}n}
\email{javier.fuentes@ugr.es}
\affiliation{Departamento de Física Teórica y del Cosmos, Universidad de Granada, Campus de Fuentenueva, E–18071 Granada, Spain}
\author{Ajdin Palavrić}
\email{ajdin.palavric@unibas.ch}
\affiliation{Department of Physics, University of Basel, Klingelbergstrasse 82, CH-4056 Basel, Switzerland}
\author{Anders Eller Thomsen}
\email{anders.thomsen@unibe.ch}
\affiliation{Department of Physics, University of Basel, Klingelbergstrasse 82, CH-4056 Basel, Switzerland}
\affiliation{Albert Einstein Center for Fundamental Physics, Institute for Theoretical Physics, University of Bern, CH-3012 Bern, Switzerland}

\date{\today}

\preprint{}

\begin{abstract}
We present a systematic method for determining the two-loop effective Lagrangian resulting from integrating out a set of heavy particles in an ultraviolet scalar theory. We prove that the matching coefficients are entirely determined from the (double-)hard region of the loop integrals and present a master formula for matching, applicable to both diagrammatic and functional approaches. We further employ functional methods to determine compact expressions for the effective Lagrangian that do not rely on any previous knowledge of its structure or symmetries. The same methods are also applicable to the computation of renormalization group equations. We demonstrate the application of the functional approach by computing the two-loop matching coefficients and renormalization group equations in a scalar toy model.
\end{abstract}

\maketitle


\section{Introduction} \label{sec:intro}

The use of Effective Field Theories (EFTs) in beyond the Standard Model (BSM) searches is ubiquitous. Given the current experimental bounds, it seems increasingly likely that there is a large gap between the electroweak scale and the next energy threshold. In this event, EFTs are the ideal tool to capture the possible low-energy effects of new physics (NP), whatever it might be. With them, we can look for subtle NP effects, opening the door to the exploration of energy scales orders of magnitude beyond what can be reached on-shell at the LHC. 

There is an ongoing community effort to automate the steps that go into EFT computations~\cite{Dawson:2022ewj,Aebischer:2023irs}. Starting with~\cite{Henning:2014wua} (see also~\cite{Fraser:1984zb,Aitchison:1984ys,Aitchison:1985pp,Aitchison:1985hu,Chan:1985ny,Chan:1986jq,Gaillard:1985uh,Cheyette:1985ue}) and further refined in~\cite{Fuentes-Martin:2016uol,Zhang:2016pja}, functional methods have been put forward as a handy way to organize one-loop EFT matching and Renormalization Group (RG) computations in BSM physics. They are used in the development of the so-called Universal One-Loop Effective Action (UOLEA)~\cite{Drozd:2015rsp,Ellis:2016enq,Ellis:2017jns,Summ:2018oko,DasBakshi:2018vni,Kramer:2019fwz,Ellis:2020ivx,Angelescu:2020yzf,Larue:2023uyv,Banerjee:2023iiv,Banerjee:2023xak,Chakrabortty:2023yke}, and 
have seen continual development~\cite{Cohen:2020fcu,Cohen:2020qvb,Fuentes-Martin:2020udw,vonGersdorff:2022kwj,vonGersdorff:2023lle}. 
It is only recently that the first fairly general tools for automated one-loop matching were introduced, one following diagrammatic amplitude matching~\cite{Carmona:2021xtq} and the other based on functional methods~\cite{Fuentes-Martin:2022jrf}. Both approaches rely on a master formula for matching that identifies the one-loop EFT action with the hard momenta region of the one-loop contributions in the underlying theory~\cite{Fuentes-Martin:2016uol,Zhang:2016pja}. 

It is now a decade since the computation of the one-loop RG equations in the Standard Model Effective Theory (SMEFT)~\cite{Jenkins:2013wua,Jenkins:2013zja,Alonso:2013hga}, and we wonder if one-loop corrections are sufficient for the current precision requirements. Indeed, certain low-energy effects are generated only at two-loop order~\cite{Bakshi:2021ofj}. Furthermore, both the strong and the top Yukawa couplings are large enough that they might generate considerable running contributions, such as in~\cite{Ardu:2021koz,Allwicher:2023aql}. On top of this, the inclusion of two-loop RG effects becomes mandatory if one wants to restore the scheme independence in one-loop matching calculations~\cite{Ciuchini:1993ks,Ciuchini:1993fk}, making them an important ingredient in the automated one-loop matching endeavor.

With this letter, we take the first step towards efficient, functional two-loop RG and matching calculations. It should come as no surprise that such a step is possible, as other variations of functional methods have been used for the calculation of the quantum effective potential~\cite{Jackiw:1974cv,Cornwall:1974vz} and the counterterms of chiral perturbation theory~\cite{Gasser:1983yg} with heat-kernel methods. Both of these calculations have also been performed at two-loop order~\cite{Ford:1992pn,Bijnens:1999hw}. What is perhaps more remarkable is that the functional formalism lets us prove a generic master formula for two-loop matching, where we directly identify the two-loop EFT action with the hard part of the ultraviolet (UV) quantum effective action. With this formula, there is no need to identify cancellations between EFT and UV contributions on a case-by-case basis. It also opens up the future possibility of automating (functional) matching beyond one-loop order. 

In this letter, we outline the extension of functional methods to two-loop order in the simpler case of scalar theories. In Section~\ref{sec:FuncMethods}, we present our functional method, with a proof to the generic matching formula supplied in Appendix~\ref{app:proof}. Next, in Section~\ref{sec:toy_example}, we demonstrate the practical application of the method by calculating the two-loop matching of a scalar toy model to its low-energy EFT and the two-loop RG equations of the resulting theory. We leave further details of the method along with extensions to fermionic degrees of freedom and gauge theories to a forthcoming paper.

\section{Functional Methods}
\label{sec:FuncMethods}

Through path-integral manipulations in a UV theory, one can determine the EFT Lagrangian describing the dynamics of the light particles at low energies compared to the heavy-particles masses. We discuss here how this can be done at the two-loop level.

\subsection{The vacuum functional and quantum effective action}

The vacuum functional, $\mathcal{W}[J]$, is the generating functional of all connected Green's functions of a theory and, therefore, contains all physical information. If we take the fields (including their conjugates) to be collectively denoted by $\eta_a(x)$ and the action by $S[\eta]$, the vacuum functional is defined by the path integral:
\begin{align} \label{eq:vac_func_def}
e^{i \hbar^{\eminus 1} \mathcal{W}[J]} = \int \mathcal{D} \eta \, e^{i\hbar^{\eminus 1}  \left( S[\eta] + J_I \eta_I \right)}\,,
\end{align} 
where the subindices with capital Latin letters $ I, J,\ldots $ correspond to DeWitt notation, where the spacetime dependence is included as part of the label, e.g. $I=(x, a)$. Thus, the contraction of repeated indices denotes not only an implicit summation in the field labels but also an integration over spacetime.\footnote{In the case at hand, $J_I \eta_I = \int_x\, J_a(x) \eta_a(x)$ with $\int_x \; \equiv \int\; \dd^d x$.} 
 
The path integral can be evaluated perturbatively, using a saddle-point approximation around the classical background-field configuration $ \overline{\eta} $, which is the solution to the tree-level equations of motion (EOMs), that is 
\begin{align}
\dfrac{\delta S^{(0)}}{\delta \eta_I}[\overline{\eta}] + J_I = 0\,,
\end{align}
where the superindex denotes loop order, with $(0)$ indicating tree level. We parameterize the expansion of the action around the background-field configuration as 
\begin{align} \label{eq:action_expansion}
S[\eta + \overline{\eta}] &\!= \!\sum_{\ell=0}^{\infty} \hbar^\ell \Big[\overline{S}^{(\ell)}\!\!+\overline{\mathcal{A}}^{(\ell)}_I \!\eta_I \!+ \overline{\mathcal{B}}^{(\ell)}_{IJ} \dfrac{\eta_I \eta_J}{2}\! + \overline{\mathcal{C}}^{(\ell)}_{IJK} \dfrac{\eta_I \eta_J \eta_K}{3!}\nonumber\\
&\quad+\overline{\mathcal{D}}^{(\ell)}_{IJKL} \dfrac{\eta_I \eta_J \eta_K \eta_L}{4!} +\ord{\eta^5} \Big]\,, 
\end{align}
with $\overline{\mathcal{A}}^{(0)}_I = -J_I$, and the bar being shorthand for exclusive dependence on $ \overline{\eta} $, e.g. $\overline{S}\equiv S[\overline{\eta}]$. 
It is convention to denote the inverse dressed propagator of the quantum field (also known as the fluctuation operator) by $\mathcal{Q}_{IJ} \equiv \mathcal{B}^{(0)}_{IJ} $. For renormalizable theories, all higher-loop vertices $\mathcal{A}_I^{(\ell)}, \mathcal{B}_{IJ}^{(\ell)},\ldots $ for $ \ell \geq 1 $ stem from the counterterms of the renormalized action; however, this could be different if our UV theory is itself an EFT.

Using the defining relation of the vacuum functional given by~(\ref{eq:vac_func_def}) together with the expansion of the action around the background field, the vacuum functional can be written in the form
\begin{widetext}
\begin{align}\label{eq:generating_functional}
\mathcal{W}[J] &= \overline{S}^{(0)}\!+ J_I \overline{\eta}_I + \hbar\,\overline{S}^{(1)}
\!+ \frac{i\hbar}{2} \big(\log \overline{\mathcal{Q}} \big)_{II}\!+ \hbar^2\,\overline{S}^{(2)} 
+\frac{i\hbar^2}{2}\overline{\mathcal Q}^{\eminus1}_{IJ} \overline{\mathcal{B}}^{(1)}_{JI}
+ \frac{\hbar^2}{12} \overline{\mathcal{C}}^{(0)}_{IJK}\overline{\mathcal Q}^{\eminus1}_{IL} \overline{\mathcal Q}^{\eminus1}_{JM} \overline{\mathcal Q}^{\eminus1}_{KN} \overline{\mathcal{C}}^{(0)}_{LMN} \nonumber\\
&\quad -\frac{\hbar^2}{8} \overline{\mathcal Q}^{\eminus1}_{IJ}\, \overline{\mathcal{D}}^{(0)}_{IJKL} \overline{\mathcal Q}^{\eminus1}_{KL} -\frac{\hbar^2}{2} \left( \overline{\mathcal{A}}^{(1)}_K+ \frac{i}{2} \overline{\mathcal Q}^{\eminus1}_{IJ} \overline{\mathcal{C}}^{(0)}_{IJK} \right) \overline{\mathcal Q}^{\eminus1}_{KL} \left( \overline{\mathcal{A}}^{(1)}_L +\frac{i}{2}\overline{\mathcal{C}}^{(0)}_{LMN} \overline{\mathcal Q}^{\eminus1}_{MN}\right)
+\ord{\hbar^3}
\,.
\end{align}    
\end{widetext} 
The expansion of the vacuum functional reproduces the usual loop expansion (formally in $\hbar$). Therefore, the terms of $\ord{\hbar^2}$ in~(\ref{eq:generating_functional}) correspond to the two-loop topologies, while the terms of higher order are dropped.

\begin{figure*}[t]
    \includegraphics[width=.9\textwidth]{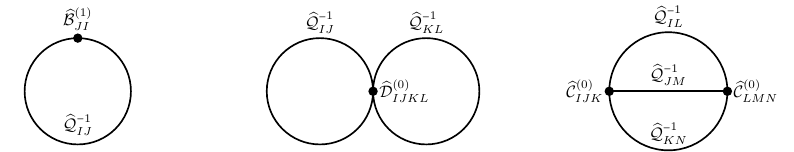}
     \caption{Graphical representation of the $\mathcal{O}(\hbar^2)$ terms appearing in the effective action in~(\ref{eq:eff_action_two-loop}).}
     \label{fig:eff_action}
\end{figure*}

It is convenient to relate this functional to the quantum effective action, $\Gamma$, as we will ultimately relate $\Gamma$ to the EFT action. The quantum effective action is the generating functional of all one-particle-irreducible (1PI) Green's functions of the theory and is defined by the Legendre transform of $\mathcal{W}$:
\begin{align} \label{eq:eff_action_def}
\Gamma[\hat{\eta}] &= \mathcal{W}[J] - J_I \hat{\eta}_I\,, &
\hat{\eta}_I &\equiv \frac{\delta\mathcal{W}}{\delta J_I}\,.
\end{align}
The background field $ \hat{\eta} $ and the sources then satisfy the quantum EOM
    \begin{align}
    \dfrac{\delta \Gamma}{\delta \eta_I}[\hat{\eta}] + J_I = 0\,.
    \end{align}
Plugging this back into definition~\eqref{eq:eff_action_def} along with the saddle-point approximation of $\mathcal{W}$, we obtain 
\begin{align} \label{eq:eff_action_two-loop}
    \Gamma[\hat{\eta}] &=  \widehat{S}^{(0)} + \hbar\,\widehat{S}^{(1)} + \dfrac{i\hbar}{2} \big(\! \log \widehat{\mathcal{Q}} \big)_{II} + \hbar^2 \widehat{S}^{(2)}\nonumber\\
    &\quad + \dfrac{i\hbar^2}{2} \widehat{\mathcal{Q}}_{IJ}^{\eminus1} \widehat{\mathcal{B}}_{JI}^{(1)} - \dfrac{\hbar^2}{8} \widehat{\mathcal{Q}}_{IJ}^{\eminus1}\, \widehat{\mathcal{D}}^{(0)}_{IJKL}  \widehat{\mathcal{Q}}_{KL}^{\eminus1} \nonumber\\ 
    &\quad + \dfrac{\hbar^2}{12}\,\widehat{\mathcal{C}}^{(0)}_{IJK} \widehat{\mathcal{Q}}_{IL}^{\eminus1} \widehat{\mathcal{Q}}_{JM}^{\eminus1} \widehat{\mathcal{Q}}_{KN}^{\eminus1} \widehat{\mathcal{C}}^{(0)}_{LMN} +\ord{\hbar^3}\,,
\end{align}
where, analogously to the bar notation, the hat is used as shorthand for exclusive dependence on $\hat\eta$, e.g. $\widehat{\mathcal{Q}}_{IJ} \equiv \mathcal{Q}_{IJ}[\hat\eta]$. This is an extension of the well-known expression for the effective action at one-loop order, where all one-loop contributions are contained in a functional (super)trace. Interestingly, the terms in the expression above can be understood as vacuum graphs with $\widehat{\mathcal{Q}}^{\eminus1}$ acting as a quantum-field propagator, dressed with arbitrary insertions of the background fields $\hat\eta$, and $\mathcal{B}^{(1)}$, $\mathcal{C}^{(0)}$ and $\mathcal{D}^{(0)}$ corresponding to two-, three-, and four-point quantum field interactions (again in the presence of background fields). This is represented in Figure~\ref{fig:eff_action}. Each closed loop in these dressed graphs can be associated to an integration over a loop momentum, in a similar manner to traditional Feynman graphs.

In general, evaluating the effective action~\eqref{eq:eff_action_two-loop} can be very complicated because of the need for inverting $\widehat{\mathcal{Q}}$ and evaluating $\widehat{\mathcal{C}}$ at different spacetime points. Nonetheless, when all loop momenta are restricted to a hard region,\footnote{The method of regions~\cite{Beneke:1997zp,Jantzen:2011nz} describes how loop integrals in dimensional regularization can be decomposed in momentum regions by expanding the integrand according to each region and integrating over the full domain (see Appendix~\ref{app:exp_by_regions}). The relevant regions for this discussion are hard and soft, with the loop momentum $k$ satisfying $k\gtrsim \Lambda$ (with $\Lambda$ being a heavy scale) and $k\ll\Lambda$, respectively.} these quantities can be evaluated directly in terms of an operator-product expansion around the hard scale. As we will now discuss, the hard-momenta region of $\Gamma$ is all that is needed for EFT matching and RG evolution, also at two-loop order.

\subsection{Master formula for EFT matching}
\label{sec:Master}

Let us consider a weakly-coupled UV theory, $ S_\sscript{\UV}[\eta] $, where $ \eta_I= (\Phi_\alpha,\, \phi_i) $ denotes the collection of all fields, heavy and light, respectively.  We seek to determine an EFT action  $ S_\EFT[\phi] $ that reproduces the physics of the full theory at energies much below the masses of the heavy fields, which we assume to lie around a generic scale $\Lambda$. 

In off-shell matching computations, we begin with an even stronger requirement for the EFT matching condition: the EFT should reproduce all low-energy Green's functions of the full theory. This stronger requirement lets us consider the generating functional of the theories rather than the $S$-matrix. 
Thus, our aim is to determine $ S_\EFT[\phi] $ such that 
\begin{align} \label{eq:vac_func_matching}
\mathcal{W}_\EFT[J_\phi] = \mathcal{W}_\UV\big[J_\Phi= 0,\, J_\phi \big]\,.
\end{align}  
That is, we enforce equality of all connected Green's functions involving the light fields. Each side of~\eqref{eq:vac_func_matching} is to be understood in terms of power series in $ 1/\Lambda $.

To make matters simpler, we proceed with a Legendre transformation of the light-field sources in order to frame the matching condition in terms of the quantum effective actions: 
\begin{align} \label{eq:matching_condition}
\Gamma_\EFT[ \hat{\phi}] &= \Gamma_\UV\big[ \widehat{\Phi}[\hat{\phi}],\, \hat{\phi}\big]\,,&  
\dfrac{\delta \Gamma_\sscript{UV}}{\delta \Phi_\alpha} \big[\widehat{\Phi}[\hat{\phi}],\, \hat{\phi}\big]=0\,.
\end{align}
The heavy fields $ \widehat{\Phi}[\hat{\phi}] $ are solutions to the quantum EOMs in the presence of the light background fields. In diagrammatic terms, condition~\eqref{eq:matching_condition} equates all one-light-particle-irreducible (1LPI) Green's functions of the UV theory with the 1PI Green's functions of the EFT. 
It was demonstrated in~\cite{Fuentes-Martin:2016uol,Zhang:2016pja} that, at one-loop order, there is a cancellation between the loop contributions in the EFT and the soft-region of the loops in the UV theory. This enables a very direct computation of $ S^{(1)}_\EFT[\phi] $ in terms of the hard region of the UV quantum effective action. 

We are now ready to generalize these considerations and present a master formula for perturbative EFT matching at multi-loop order: the off-shell EFT action is determined by  
    \begin{align} \label{eq:gen_matching_formula}
    S_\EFT &= \Gamma_\UV\big[ \widehat{\Phi}[\hat{\phi}],\, \hat{\phi}\big] \Big|_\mathrm{hard}\,,&
    \dfrac{\delta \Gamma_\sscript{UV}\big|_\mathrm{hard} }{\delta \Phi_\alpha} \big[ \widehat{\Phi}[\hat{\phi}],\, \hat{\phi} \big]&=0\,.
    \end{align}
Here the `hard' part is taken to include all contributions without \emph{any} soft loops. It includes tree-level contributions as well as loops where \emph{all} loop momenta are hard. In many ways, this is an intuitive leap: the hard, local part of the UV theory is identified with the EFT action, while the long-distance physics is captured by loops in the EFT. Nevertheless, we have never seen an explicit formulation of this notion, much less a matching formula applicable to practical computations. A constructive proof of the matching formula~\eqref{eq:gen_matching_formula} at two-loop order is provided in Appendix~\ref{app:proof}, where we show that there is a one-to-one correspondence between (partially) soft loops in the UV and loops in the EFT. We postpone the discussion on a possible extension of this proof to higher-loop orders to a more comprehensive follow-up paper. The end result is that all loop integrals needed to perform the EFT matching (i.e. those in the hard limit) reduce to vacuum integrals, for which expressions are known up to three loops~\cite{Martin:2016bgz}. 

The matching formula~\eqref{eq:gen_matching_formula} is a $d$-dimensional off-shell relation. A complication associated to this kind of matching is that it does not produce the EFT Lagrangian directly in a four-dimensional on-shell basis. Rather, one has to apply field redefinitions to reduce the output to an on-shell basis. Likewise, the matching result may also produce EFT operators that are not present in a four-dimensional basis. As a result, one has to separate out evanescent operators and, preferably, convert the EFT action to an evanescence-free scheme~\cite{Dugan:1990df,Buras:1989xd,Herrlich:1994kh,Fuentes-Martin:2022vvu,Aebischer:2022tvz,Aebischer:2022aze,Aebischer:2022rxf}. A related consideration to be aware of beyond one-loop order is that lower-order matching coefficients may contain $\ord{\epsilon}$ contributions. These cannot be ignored, as their insertion in an EFT loop can lead to finite contributions. Therefore, one has to carefully remove the $\ord{\epsilon}$ terms and absorb them into finite coefficients at higher-loop order, similarly to what is done for evanescent contributions~\cite{Fuentes-Martin:2022vvu}.

\subsection{Functional approach for RG evolution}

The MS (or \msbar) counterterms of a theory, $S[\eta]$, can also be determined from the effective action from the observation that it must be free of UV divergences. If we use dimensional regularization to regularize the loop integrals, finiteness of the renormalized generating functional translates to the condition 
\begin{align}\label{eq:FreeOfUVPoles}
K_\epsilon\,\Gamma[\hat \eta]=0\,,
\end{align}
where $K_\epsilon$ is an operator that extracts all $1/\epsilon$ poles of UV origin. This equation establishes a relation between the MS counterterms of the theory, identified with the UV poles of $\widehat S^{(\ell)}$, and the other terms in the effective action. Denoting the MS counterterms by $\delta \widehat S^{(\ell)}_{\rm MS}$, we have
\begin{align}
\delta \widehat S^{(\ell)}_{\rm MS}=K_\epsilon\,\widehat S^{(\ell)}\,,
\end{align}
which, together with the expression of $\Gamma$ and condition~\eqref{eq:FreeOfUVPoles}, establishes a direct relation to determine $\delta \widehat S^{(\ell)}_{\rm MS}$ functionally. Other renormalization schemes of the subtraction family can also be obtained by appropriately adapting the definitions of the counterterms in the expression above. 

Restricting to the two-loop expression of the effective action in~\eqref{eq:eff_action_two-loop}, we obtain the following MS counterterms up to two-loop order:
\begin{align}\label{eq:FunctionalCTs}
\delta \widehat S^{(1)}_{\rm MS}&=-\dfrac{i}{2} K_\epsilon\big(\! \log \widehat{\mathcal{Q}} \big)_{II}\,,\nonumber\\
\delta \widehat S^{(2)}_{\rm MS}&= K_\epsilon\Big[-\frac{i}{2}\widehat{\mathcal Q}^{\eminus1}_{IJ} \widehat{\mathcal{B}}^{(1)}_{JI} +\dfrac{1}{8} \widehat{\mathcal{Q}}_{IJ}^{\eminus1} \widehat{\mathcal{D}}^{(0)}_{IJKL}  \widehat{\mathcal{Q}}_{KL}^{\eminus1} \nonumber\\ 
&\quad - \dfrac{1}{12}\,\widehat{\mathcal{C}}^{(0)}_{IJK} \widehat{\mathcal{Q}}_{IL}^{\eminus1} \widehat{\mathcal{Q}}_{JM}^{\eminus1} \widehat{\mathcal{Q}}_{KN}^{\eminus1} \widehat{\mathcal{C}}^{(0)}_{LMN}\Big]\,.
\end{align}
A convenient prescription for extracting the UV poles in theories with massless states consists in introducing a common mass, $\Lambda$, in all propagators of the loop integrals. This mass acts as a hard scale (assumed to be much larger than any other scales in the loop integrals) and serves as an infrared regulator. 
The overall UV divergence of a loop integral is identified with the part where all loop momenta are large. Hence, it is easy to show that the UV divergences can be extracted from the hard-region (defined by all loop momenta being of order $\Lambda$) of the loop integrals~\cite{Chetyrkin:1997fm}. This effectively establishes a power-counting on $\Lambda$, around which $\widehat{\mathcal{Q}}^{-1}$ can be expanded. As in the matching case, the resulting loop integrals are just vacuum integrals (in this case with a single mass $\Lambda$) for which results are known up to three loops~\cite{Martin:2016bgz}. The only drawback of this method for UV-pole extraction is that one also needs to consider spurious counterterms with positive powers of $\Lambda$. These spurious counterterms can break the symmetries of the original Lagrangian and are needed for the cancellation of subdivergences. After the counterterms have been determined, the RG equations can be readily obtained via standard techniques.

\subsection{Functional evaluation of the effective action}
\label{subsec:func_evaluations}

The functional evaluation of the effective action requires manipulating the inverse dressed propagator, $\widehat{\mathcal{Q}}$, which can be generically parametrized by
\begin{align}
\widehat{\mathcal{Q}}_{IJ} \equiv \widehat{\mathcal{Q}}_{ab}(x,y) = Q_{ab}(x,\,P_x)\,\delta(x-y)\,,
\end{align}
with $ P^\mu_x = i \partial^\mu_x $ denoting the momentum operator. Since the action is local, it is always possible to factor out a delta function. This is also the case when dealing with more complicated functional forms involving $\widehat{\mathcal{Q}}$, such as $\ln \widehat{\mathcal{Q}}$ or $\widehat{\mathcal{Q}}^{\eminus 1}$. It is well-known that the one-loop contribution to the quantum effective action, c.f.~\eqref{eq:eff_action_two-loop}, is given by the functional (super)trace
\begin{align}
\widehat\Gamma^{(1)}&=\widehat S^{(1)}+\frac{i}{2}(\ln\widehat{\mathcal{Q}})_{II}\nonumber\\
&=\widehat S^{(1)}+\frac{i}{2}\int_{x,k}\;\,[\ln Q(x,P_x-k)]_{aa}\,,
\end{align}
with $\int_{x,k}\equiv\int\, \dd^dx \int\, \dd^dk/(2\pi)^d$. This expression is a non-local function of $Q$ and is difficult to evaluate in general. However, as we are only interested in the hard region, where the loop momentum is taken to be of the order of the heavy scales, we can perform an operator-product expansion. For scalar theories, the inverse dressed propagator takes the generic form
\begin{align}
Q_{ab}(x,\,P_x)=(P_x^2-M_{a}^2)\,\delta_{ab}-U_{ab}(x, P_x)\,,
\end{align}
with $M_a$ being a possible (hard) mass, and $U_{ab}$ a generic interaction term which, as we make explicit in its argument, may involve derivatives. Denoting 
\begin{align}
\Delta^{\! \eminus 1}_{ab}(P_x,k)&\equiv(k^2-M_a^2+P_x^2-2k\cdot P_x)\,\delta_{ab}\,,\nonumber\\
X_{ab}(x,P_x,k)&\equiv U_{ab}(x,P_x-k)\,,
\end{align}
the operator-product expansion of the logarithm reads
\begin{align} \label{eq:log_expansion}
\int_{x,k}\ln Q(x,P_x-k)&= \int_{x,k} \ln\Delta^{\! \eminus1} -\int_{x,k}\;\, \sum_{n=1}^\infty\frac{1}{n}\,(\Delta X)^n\,,
\end{align}
where $(\ln\Delta^{\! \eminus 1})_{ab}=\delta_{ab}\ln(k^2-M_a^2)$ contributes to an unphysical constant that is subtracted when normalizing the path integral and the expansion for $\Delta$ takes the form 
\begin{align}\label{eq:Delta_expansion}
\Delta_{ab}(P_x,k)&=\delta_{ab}\sum_{n=0}^\infty\frac{(-P_x^2+2k\cdot P_x)^n}{(k^2-M_a^2)^{n+1}}\,.
\end{align}
Likewise, the dressed propagator, necessary for evaluating the two-loop contributions, admits the expansion
\begin{align}\label{eq:OPE_Qm1}
Q^{\eminus1}(x,P_x-k)=\sum_{n=0}^\infty(\Delta X)^n\Delta\,.
\end{align}
The hard-region evaluation guarantees that subsequent terms in all these series are further suppressed, so only a finite number of terms need to be retained to a given order in the EFT expansion. A manifestly local result is obtained only in the hard-region limit.

Locality of the action likewise ensures that delta functions can be factored out of the remaining functional objects. We write the quantum-field interactions as series in the momenta operator:\footnote{For compactness, we employ a power-like notation with underlined superscripts for the Lorentz indices, that is, we denote $B_{ab}^{(1)\,\underline{m}} P_x^{\underline{m}} \equiv B_{ab}^{(1)\, \mu_1 \ldots \mu_m} P_x^{\mu_1} \cdots  P_x^{\mu_m}$.} 
\begin{align}
\widehat{\mathcal{B}}^{(1)}_{IJ} &=\widehat{\mathcal{B}}^{(1)}_{ab}(x,y)= \sum_{m=0}^\infty B^{(1)\, \underline{m}}_{ab}(x) P_x^{\underline{m}} \,\delta(x-y)\,,\nonumber\\
\widehat{\mathcal{C}}^{(0)}_{IJK} &= \widehat{\mathcal{C}}^{(0)}_{abc}(x,y,z) \nonumber\\
&= \sum_{m,n=0}^\infty C^{\underline{m},\underline{n}}_{abc}(z) P_x^{\underline{m}} P_y^{\underline{n}}\,\delta(x-z) \delta(y-z)\,, \nonumber\\
\widehat{\mathcal{D}}^{(0)}_{IJKL} &= \widehat{\mathcal{D}}^{(0)}_{abcd}(x,y,z,w) = \sum_{m,n,r=0}^\infty D^{\underline{m},\underline{n},\underline{r}}_{abcd}(w)  \nonumber\\
&\quad \times P_x^{\underline{m}} P_y^{\underline{n}} P_z^{\underline{r}}\,\delta(x-w) \delta(y-w) \delta(z-w)\,.  
\end{align}
In most practical applications, only a small number of terms from this momentum operator expansions are present. In fact, for renormalizable theories, only the terms with at most two momenta in $\widehat{\mathcal{B}}$, one in $\widehat{\mathcal{C}}$, and none in $\widehat{\mathcal{D}}$ are nonzero. 

Having made these definitions, we can now evaluate the two-loop contributions to the effective action in~\eqref{eq:eff_action_two-loop}. We parameterize them as 
\begin{align}
\widehat{\Gamma}^{(2)} = \widehat{S}^{(2)} +  \dfrac{i}{2} G_{\mathrm{ct}} -\dfrac{1}{8} G_{\mathrm{f8}} + \dfrac{1}{12} G_\mathrm{ss}\,,
\end{align}
where the loop contributions $G_i$ are identified with the counterterm, figure-8, and sunset topologies, respectively (as depicted in Figure~\ref{fig:eff_action}). The main subtlety in the functional evaluation of the two-loop effective action is related to the sunset topology, where it is necessary to power expand one of the two vertices around the location of the other. We then obtain integral formulas for the functional contractions and, using a momentum-space representation for the delta functions, we find the expressions 
\begin{align}\label{eq:two_loop_topologies}
G_{\mathrm{ct}}&=\sum_m \! \int_{x,k} B^{(1)\, \underline{m}}_{ab}(x)\big[(P_x-k)^{\underline{m}}\,Q_{ba}^{\eminus1}(x,\,P_x -k )\big]\,,\nonumber\\[5pt]
G_{\mathrm{f8}}&=\sum_{m,n,r}(\eminus 1)^{m+n+r} \!\! \int_{x,k,\ell} D^{\underline{m},\underline{n},\underline{r}}_{abcd}(x)\,k^{\underline{m}} \,\ell^{\underline{r}}\nonumber\\*
&\quad \times \big[(P_x-k)^{\underline{n}}\,Q_{ba}^{\eminus1}(x,\,P_x -k ) \big] \big[Q_{dc}^{\eminus1}(x,\,P_x - \ell)\big]\,,\nonumber\\[5pt]
G_\mathrm{ss} &= \!\! \!\! \sum_{m^{(\prime)},\,n^{(\prime)},\,s} \!\!\! \!\!\!(\eminus 1)^{m+n+m^\prime+n^\prime}\dfrac{i^s}{s!} \int_{x,k,l} C^{\underline{m}, \underline{n}}_{abc}(x) \partial_x^{\underline{s}}\,C^{\underline{m}', \underline{n}'}_{def}(x)\nonumber\\*
&\quad \times\big[\partial_k^{\underline{s}}\,Q_{cf}^{\eminus 1}(x,\,P_x +k + \ell) \big]\, k^{\underline{m}'} \ell^{\underline{n}'}\nonumber\\*
&\quad \times\big[(P_x-k)^{\underline{m}}\,Q_{ad}^{\eminus 1}(x,\,P_x-k) \big]   \nonumber\\*
&\quad \times\big[(P_x-\ell)^{\underline{n}}\,Q_{be}^{\eminus 1}(x,\, P_x-\ell ) \big]   \,,
\end{align}  
with $\int_{x,k,\ell}\equiv\int\, \dd^dx \int\, \dd^dk/(2\pi)^d\int\, \dd^d\ell/(2\pi)^d$. In the equations above, any $P_x$ acting to the rightmost of a bracket yields a null contribution. This would not be the case anymore in the gauge non-singlet scenario, where the derivative in $P_x$ would be promoted to a covariant derivative. A manifestly covariant generalization of these expressions will be presented in a follow-up paper. While the sum in $s$ in the last expression runs to infinity, only a finite number of terms need to be retained at a given order in the EFT counting when considering the hard-momentum limit. In particular, only terms up to $s=4$ contributes at EFT dimension six. 

\section{A toy-model example}
\label{sec:toy_example}

We illustrate the functional method described in the previous section with a concrete example: a toy-model, consisting of one heavy and one light real scalar fields,  $\Phi$ and  $\phi$, respectively. For simplicity, we assume that the theory possesses a $\mathbb{Z}_2^\phi\times \mathbb{Z}_2^\Phi$ symmetry, with $\mathbb{Z}_2^\Phi$ softly-broken by a trilinear term with coupling $ \kappa \ll M$. This soft-breaking term is included to allow for a non-trivial quantum EOM of the heavy-field, cf.~\eqref{eq:gen_matching_formula}. The UV Lagrangian of this theory is given by
\begin{align}\label{eq:scalar_theory_UV_lagr}
\mathcal{L}_\UV&=\frac{1}{2}(\partial_\mu \phi)^2+\frac{1}{2}(\partial_\mu\Phi)^2-\frac{1}{2}m_\phi^2\phi^2-\frac{1}{2}M^2\Phi^2-\frac{\lambda_\phi}{4!}\phi^4\nonumber\\
&\quad-\frac{\lambda_\Phi}{4!}\Phi^4-\frac{\lambda_{\Phi\phi}}{4}\Phi^2\phi^2-\frac{\kappa}{3!}\Phi^3+\mathcal{L}^{\rm ct}_\UV\,,
\end{align}
with the corresponding UV counterterms $\mathcal{L}^{\rm ct}_{\rm UV}$ defined as
\begin{align}\label{eq:scalar_theory_UV_counter}
\mathcal{L}^{\rm ct}_\UV&=\!\frac{\delta_\phi}{2}(\partial_\mu\phi)^2\!+\frac{\delta_\Phi}{2}(\partial_\mu\Phi)^2-\frac{\delta_{m_\phi}}{2}\phi^2-\frac{\delta_M}{2}\Phi^2-\frac{\delta_{\lambda_\phi}}{4!}\phi^4\nonumber\\
&\quad-\frac{\delta_{\lambda_\Phi}}{4!}\Phi^4-\frac{\delta_{\lambda_{\Phi\phi}}}{4}\Phi^2\phi^2-\frac{\delta_\kappa}{3!}\Phi^3-\frac{\delta_{\kappa^\prime}}{2}\Phi\phi^2-\delta_{\kappa_T}\Phi\,,
\end{align}
where we anticipate that new $\mathbb{Z}_2^\Phi$-breaking interactions are generated radiatively and need to be renormalized.\footnote{The trilinear coupling $\Phi \phi^2$ is generated radiatively in the RG and should properly be included. As we are not attempting to build a realistic model, we simply ignore it.}

\subsection{One-loop UV counterterms}

Our calculations are done in a tadpole-free \msbar scheme (for both UV theory and the EFT), such that the $ \Phi $ tadpole is removed with a finite counterterm. Given this scheme choice, the determination of the finite parts of the two-loop matching conditions requires only the calculation of the one-loop counterterms. From~\eqref{eq:FunctionalCTs} and with the functional evaluation described in Section~\ref{subsec:func_evaluations}, it follows that\footnote{Not only the $1/\epsilon_\UV$ pole but also the finite pieces are retained for $\Phi$ tadpole counterterms. Functional techniques are still appropriate for evaluating the finite tadpole, as it is independent of external momenta.} 
\begin{align}
\mathcal{L}^{\rm ct\,(1)}_\UV &=\frac{i}{2}\,K_\epsilon \! \int_k\;\; \sum_{n=1}^\infty\frac{1}{n}[(\Delta X)^n]_{aa}\,,
\end{align}
where $a=\Phi, \phi$. The functional objects relevant for this calculation are given by
\begin{align}\label{eq:DeltaUs_example}
X_{\Phi\Phi}(x,P_x,k)&= \frac{\lambda_\Phi}{2}\Phi^2+\frac{\lambda_{\Phi\phi}}{2}\phi^2+\kappa\Phi\,,\nonumber\\
X_{\phi\phi}(x,P_x,k)&= \frac{\lambda_\phi}{2}\phi^2+\frac{\lambda_{\Phi\phi}}{2}\Phi^2\,,\nonumber\\
X_{\Phi\phi}(x,P_x,k)&= X_{\phi\Phi}(x,P_x,k)=\lambda_{\Phi\phi}\,\Phi\phi\,,
\end{align}
and the expansion of $\Delta$ in~\eqref{eq:Delta_expansion}. As the UV theory in our example contains no massless states, no IR divergences appear and the UV divergences are readily obtained. Only the terms with $n\leq 2$ in the sum above contribute to the UV divergences. Denoting $\delta_i\equiv\frac{\hbar}{16\pi^2}\,\delta_i^{(1)}+\ord{\hbar^2}$, we find
\begin{align}
\delta_\phi^{(1)}&=\delta_\Phi^{(1)}=0\,,&
\delta_{\kappa}^{(1)}&=\frac{3\lambda_\Phi \kappa}{2\epsilon}\,,\nonumber\\
\delta_{m_\phi}^{(1)}&= \frac{\lambda_\phi m_\phi^2+ \lambda_{\Phi\phi} M^2}{2\epsilon}\,,&
\delta_{\lambda_\Phi}^{(1)}&=\frac{3\lambda_\Phi^2+3 \lambda_{\Phi\phi}^2}{2\epsilon}\,,\nonumber\\
\delta_M^{(1)}&=\frac{\lambda_\Phi M^2+ \lambda_{\Phi\phi} m_\phi^2+\kappa^2}{2\epsilon}\,,&
\delta_{\lambda_\phi}^{(1)}&=\frac{3\lambda_\phi^2+3 \lambda_{\Phi\phi}^2}{2\epsilon}\,,\nonumber\\
\delta_{\lambda_{\phi\Phi}}^{(1)}&=\frac{\lambda_\phi\lambda_{\Phi\phi} + \lambda_\Phi\lambda_{\Phi\phi}+4\lambda_{\Phi\phi}^2}{2\epsilon}\,,&
\delta_{\kappa^\prime}^{(1)}&= \frac{\lambda_{\Phi\phi}\kappa}{2\epsilon}\,,\nonumber\\
\delta_{\kappa_T}^{(1)}&=\frac{M^2\kappa}{2}\left(\frac{1}{\epsilon}-\overline\ln\,M^2+1\right)\,,
\end{align}
where $\overline\ln\,M^2\equiv\ln M^2/\bar\mu^2$ and $\bar\mu$ is the \msbar renormalization scale. The value of these counterterms has been cross-checked against the RG functions obtained with RGBeta~\cite{Thomsen:2021ncy}.

\subsection{EFT matching at two-loop order}

For the present example, it is easy to see that there is no tree-level contribution to the EFT action, as the Lagrangian has no linear dependence on the heavy field $\Phi$. The one-loop part of the EFT Lagrangian is calculated as usual from the hard region of the functional (super)trace which, after doing the expansion~\eqref{eq:log_expansion}, yields 
\begin{align}
\mathcal{L}^{(1)}_\EFT&=-\frac{i}{2}\int_k\;\;\sum_{n=1}^\infty\frac{1}{n}[(\Delta X)^n]_{aa}\bigg|_{\rm hard}\,,
\end{align}
with $\Delta$ and $X$ provided in~\eqref{eq:Delta_expansion} and~\eqref{eq:DeltaUs_example}, respectively. In contrast with the counterterm evaluation, we are now interested in the hard-momentum region defined by the relation $k\gtrsim M\gg m_\phi$, with $k$ being the loop momentum. In practice, this implies that, for this calculation, the mass term in $\Delta^{\! \eminus1}_{\phi\phi}$ needs to be power-expanded before loop integration. That is, 
\begin{align}\label{eq:Delta_phiphi}
\Delta^{\! \eminus 1}_{\phi\phi}(P_x,k)\big|_{\rm hard}=\sum_{n=0}^\infty\frac{(-P_x^2+2k\cdot P_x+m_\phi^2)^n}{k^{2(n+1)}}\,.
\end{align}

One difference to keep in mind with respect to the usual one-loop matching evaluations is the need to retain $\ord{\epsilon}$ terms. As described in Section~\ref{sec:Master}, these terms can be shifted into the two-loop matching coefficients with an appropriate modification of the renormalization conditions. The one-loop EFT Lagrangian resulting from the functional trace is a function of both heavy and light fields. The former are removed by means of the heavy-field EOMs defined in~\eqref{eq:gen_matching_formula}. In our case, we find that
\begin{align}
\Phi= -\frac{\hbar}{16\pi^2} \frac{\lambda_{\Phi\phi}\kappa}{2M^2}\, \overline\ln\,M^2\, \frac{\phi^2}{2} +\mathcal{O}(M^{ \eminus 4},\hbar^2)\,.
\end{align} 
The contributions from replacing $\Phi$ by its EOM yield two-loop effects when inserted back into the one-loop (and tree-level) EFT Lagrangian. No terms of two-loop order are needed in our toy-model calculation, as these would start contributing only at the three-loop level. 

The two-loop EFT Lagrangian follows directly from the evaluation of expressions~\eqref{eq:two_loop_topologies} in the double-hard region, defined by $k,\ell \gtrsim M\gg m$. In this example, only terms without powers of $P_{x,y,z}$ appear in $\widehat{\mathcal{B}}^{(1)}$, $\widehat{\mathcal{C}}$, and $\widehat{\mathcal{D}}$, and thus, only the terms with $m^{(\prime)}=n^{(\prime)}=r=0$ in~\eqref{eq:two_loop_topologies} contribute. We have
\begin{align}
B^{(1)}_{\Phi\Phi}&=-\delta_M^{(1)}-\frac{\delta_{\lambda_{\Phi\phi}}^{(1)}}{2}\phi^2\,,&
\!\!\!B^{(1)}_{\phi\phi}&=-\delta_{m_\phi}^{(1)}-\frac{\delta_{\lambda_\phi}^{(1)}}{2}\phi^2\,,\nonumber\\
B^{(1)}_{\Phi\phi}&=-\delta_{\kappa_{\Phi\phi}}^{(1)}\phi\,,\nonumber\\[10pt]
C_{\Phi\Phi\Phi}&=-\kappa\,,&
\!\!\!C_{\phi\phi\phi}&=-\lambda_{\phi}\phi\,,\nonumber\\
C_{\Phi\Phi\phi}&=-\lambda_{\Phi\phi}\phi\,,&
\!\!\!C_{\Phi\phi\phi}&=0\,,\nonumber\\[10pt]
D_{\Phi\Phi\Phi\Phi}&=-\lambda_\Phi\,,&
\!\!\!D_{\phi\phi\phi\phi}&=-\lambda_\phi\,,\nonumber\\
D_{\Phi\Phi\phi\phi}&=-\lambda_{\Phi\phi}\,,&
\!\!\!D_{\Phi\Phi\Phi\phi}&=D_{\Phi\phi\phi\phi}=0\,,
\end{align}
where we dropped terms containing $\Phi$, as these contribute only at higher loop orders. 
The final (off-shell) EFT Lagrangian is obtained by combining the results from~\eqref{eq:two_loop_topologies} with the contributions from the $\Phi$-field EOM and the two-loop shift from removing the one-loop $\ord{\epsilon}$ terms. The resulting EFT Lagrangian can be written in terms of an on-shell operator basis via appropriate $\phi$-field redefinitions. For concreteness, we present here this final on-shell result:
\begin{align}\label{eq:starting_eft_lagrangian}
\mathcal{L}_\EFT&=\frac{1}{2}(\partial_\mu\phi)^2-\frac{1}{2}m^2\phi^2-\frac{\lambda}{4!}\phi^4-\frac{c_6}{6!}\phi^6 +\mathcal{L}_\EFT^\mathrm{ct}\,,
\end{align}
with the two-loop matching conditions between the EFT and UV Lagrangians taking the form 
\begin{widetext}
\begin{align}
m^2& =m_\phi^2- \frac{\lambda_{\Phi\phi}}{32 \pi^2} \left(1- \overline\ln\, M^2\right) M^2 +\frac{1}{(16\pi^2)^2} \left[ \left(5- 4\,\overline\ln\, M^2+ \overline\ln^{\,2} M^2 \right) \frac{\lambda_{\Phi\phi}^2}{2} M^2- \left( \overline\ln\,M^2- \overline\ln^2M^2 \right) \frac{\lambda_\Phi \lambda_{\Phi\phi}}{4} M^2\right. \nonumber\\
&\quad\left. +\left(1- 2\sqrt{3}\, \mathrm{Cl}_2- 2\,\overline\ln\, M^2+ \overline\ln^2M^2 \right) \frac{\lambda_{\Phi\phi}}{8} \kappa^2 -\left(\frac{11}{4}+ \overline\ln\,M^2+ \overline\ln^{\,2} M^2 \right) \frac{\lambda_{\Phi\phi}^2}{4} m_\phi^2+ \left(5+ 6\,\overline\ln\, M^2 \right) \frac{\lambda_{\Phi\phi}^2 m_\phi^4}{36M^2} \right]\,,\nonumber\\[5pt]
\lambda&= \lambda_\phi+ \frac{1}{16\pi^2} \left[\frac{3\lambda_{\Phi\phi}^2}{2}\, \overline\ln\,M^2 -\frac{\lambda_{\Phi \phi}^2m_\phi^2}{3M^2} \right]+ \frac{1}{(16 \pi^2)^2} \left[ \left(19-37\,\overline\ln\,M^2+ 18\,\overline\ln^2\,M^2 \right) \frac{\lambda_{\Phi\phi}^3}{6}-\left(11+ 26\, \overline\ln\, M^2 \right) \frac{\lambda_{\Phi\phi}^3 m_\phi^2}{6M^2} \right. \nonumber\\
&\quad\left.-\left(35+ 20\,\overline\ln \,M^2+ 12\, \overline\ln^2 M^2 \right) \frac{\lambda_{\Phi\phi}^2 \lambda_\phi}{8} +\left(1 + \overline\ln\, M^2\right) \frac{4\lambda_{\Phi\phi}^2 \lambda_\phi m_\phi^2}{3M^2}- \left(1- \overline\ln\, M^2- \overline\ln^2\, M^2 \right) \frac{3\lambda_\Phi\lambda_{\Phi\phi}^2}{4} \right.\nonumber\\
&\quad\left. -\left(1+ \overline\ln\,M^2 \right) \frac{\lambda_\Phi \lambda_{\Phi\phi}^2m_\phi^2}{6M^2}- \left(1- \overline\ln\, M^2+ \overline\ln^2\, M^2\right) \frac{3\kappa^2 \lambda_{\Phi\phi}^2}{4M^2} \right]\,,\nonumber\\[5pt]
c_6&=\frac{1}{16\pi^2} \left[\frac{15 \lambda_{\Phi\phi}^3}{2M^2}- \frac{5 \lambda_{\Phi\phi}^2 \lambda_\phi}{3M^2} \right]+ \frac{1}{(16\pi^2)^2} \left[ -\bigg( 18-17\, \overline\ln\,M^2 \bigg) \frac{5\lambda_{\Phi\phi}^4}{2M^2}- \bigg(47+ 62\,\overline\ln\, M^2 \bigg) \frac{5\lambda_{\Phi\phi}^3 \lambda_\phi}{6M^2} \right.\nonumber\\
&\quad \left.+ \bigg(13+ 10\,\overline\ln\,M^2 \bigg) \frac{5\lambda_{\Phi\phi}^2 \lambda_\phi^2}{6M^2}+ \left(1+ 2\,\overline\ln\, M^2 \right) \frac{15\lambda_\Phi \lambda_{\Phi\phi}^3}{4M^2}- \left(1+ \overline\ln\,M^2 \right) \frac{5\lambda_\Phi \lambda_{\Phi\phi}^2 \lambda_\phi}{6M^2}\right]\,,
\end{align}
\end{widetext}
where $\mathrm{Cl}_2(x)$ is the Clausen function of order 2 with $\mathrm{Cl}_2\equiv \mathrm{Cl}_2(2\pi/3)\approx0.6766277$. To our knowledge, this is the first two-loop matching computation performed with functional methods. 
As a crosscheck of our result, we verified that all single- and double-logarithmic contributions are consistent with the RG functions in both the UV theory\footnote{The two-loop RG functions in the UV theory necessary for this crosscheck can easily be determined using \texttt{RGBeta}~\cite{Thomsen:2021ncy}.} and the EFT such as to ensure matching scale independence. The determination of the RG functions in the EFT using functional methods is discussed in the next section.

\subsection{Renormalization group functions at two-loop order}

The EFT counterterms are calculated functionally using expressions~\eqref{eq:FunctionalCTs}. Analogously to the UV counterterms, the EFT counterterms at one-loop order are given by
\begin{align}
\mathcal{L}^{\rm ct\,(1)}_\EFT&= \frac{i}{2}\,K_\epsilon\!\int_k\;\;\sum_{n=1}^\infty\frac{1}{n}[(\Delta_\EFT X_\EFT)^n]_{aa}\,,
\end{align}
with $\Delta_\EFT$ as in~\eqref{eq:Delta_expansion} and $X_\EFT$ being
\begin{align}
X_\EFT(x,P_x,k)&=\frac{\lambda}{2}\phi^2+\frac{c_6}{4!}\phi^4\,.
\end{align}
The counterterm Lagrangian from the expression above is obtained in an off-shelf basis, and can be reduced to an on-shell Lagrangian by appropriate field redefinitions. Parameterizing the on-shell EFT counterterms by
\begin{align}
\mathcal{L}_\EFT^\mathrm{ct}&=\frac{1}{2}\delta_\phi(\partial_\mu\phi)^2-\frac{1}{2}\delta_{m^2}\phi^2-\frac{\delta_\lambda}{4!}\phi^4-\frac{\delta_{c_6}}{6!}\phi^6\,,
\end{align}
and separating the couplings by loop order and power of the $\epsilon$-pole as
\begin{align}
\delta_i= \frac{\hbar}{16\pi^2} \frac{\delta_i^{(1)}}{\epsilon} +\frac{\hbar^2}{(16\pi^2)^2} \left[ \frac{\delta_{i,1}^{(2)}}{\epsilon} +\frac{\delta_{i,2}^{(2)} }{\epsilon^2} \right]+\ord{\hbar^3}\,,
\end{align}  
we find
\begin{align}\label{eq:one-loop_cts}
\delta^{(1)}_\phi&=0\,,&
\delta^{(1)}_{m^2}&=\frac{m^2\lambda}{2}\,,\nonumber\\*
\delta^{(1)}_\lambda&=\frac{3\lambda^2+m^2c_6}{2}\,,&
\delta_{c_6}^{(1)}&=\frac{15\lambda c_6}{2}\,,
\end{align}
at one-loop order.

The two-loop part of the counterterms is then obtained from the second line of~\eqref{eq:FunctionalCTs}. Once more, only terms with no powers of $P_{x,y,z}$ in $\widehat{\mathcal{B}}^{(1)}_\EFT$, $\widehat{\mathcal{C}}_\EFT$, and $\widehat{\mathcal{D}}_\EFT$ are present, and we have that
\begin{align}
B_\EFT^{(1)}&=-\frac{1}{16\pi^2} \frac{1}{\epsilon}\left[\delta^{(1)}_{m^2}+\frac{\delta_\lambda^{(1)}}{2}\phi^2+\frac{\delta_{c_6}^{(1)}}{4!}\phi^4\right]\,,\nonumber\\
C_\EFT&=-\lambda\phi-\frac{c_6}{3!}\phi^3\,,\qquad
D_\EFT=-\lambda-\frac{c_6}{2}\phi^2\,,
\end{align}
where we used that $\delta_\phi^{(1)}=0$ in this example. Inserting these operators into the expressions in~\eqref{eq:two_loop_topologies} (where, again, only the terms with $m^{(\prime)}=n^{(\prime)}=r=0$ contribute) together with the expansion of $Q^{-1}$ in~\eqref{eq:OPE_Qm1} and evaluating the loop integrals, we find the two-loop contributions to the counterterms. As before, those are obtained in an off-shell basis, which reduces to the on-shell result 
\begin{align}\label{eq:two-loop_cts}
\delta_{m^2,1}^{(2)}&=-\frac{\lambda^2m^2}{4}\,,&
\delta_{m^2,2}^{(2)}&=\frac{4m^2\lambda^2+m^4 c_6}{8}\,,\nonumber\\
\delta_{\lambda,1}^{(2)}&=-\frac{9\lambda^3+5m^2\lambda c_6}{6}\,,&
\delta_{\lambda,2}^{(2)}&=\frac{9\lambda^3+11m^2\lambda c_6}{4}\,,\nonumber\\
\delta_{c_6,1}^{(2)}&=-\frac{215\lambda^2c_6}{12}\,,&
\delta_{c_6,2}^{(2)}&=\frac{135\lambda^2c_6}{4}\,,\nonumber\\
\delta_{\phi,1}^{(2)}&=-\frac{\lambda^2}{24}\,,&
\delta_{\phi,2}^{(2)}&=0\,,
\end{align}
after appropriate field redefinitions. As a cross-check, we have verified that these counterterms satisfy the consistency conditions on the double-poles necessary for finite RG functions (see, e.g.~\cite{Herren:2021yur}).

Having obtained the two-loop counterterms, we readily determine the anomalous dimension $\gamma_\phi$ of the field $\phi$ along with the beta functions for $m^2$, $\lambda$ and $c_6$ couplings:
\begin{align}\label{eq:beta_functions}
\gamma_\phi&=\frac{1}{2}\frac{\dd\ln(1+\delta_\phi)}{\dd \ln\mu} = -\frac{\partial\delta_{\phi,1}^{(2)}}{\partial\lambda}=\frac{1}{(16\pi^2)^2}\frac{\lambda^2}{12}\,,\nonumber\\
\beta_{m^2}&=\frac{m^2\lambda}{16\pi^2}-\frac{5}{6}\frac{m^2\lambda^2}{(16\pi^2)^2}\,,\nonumber\\
\beta_\lambda&=\frac{3\lambda^2+m^2 c_6}{16\pi^2}-\frac{1}{3}\frac{17\lambda^3+10\lambda m^2 c_6}{(16\pi^2)^2},\nonumber\\
\beta_{c_6}&=\frac{15\lambda c_6}{16\pi^2}-\frac{427}{6}\frac{\lambda^2 c_6}{(16\pi^2)^2}\,.
\end{align}
Those terms of the beta functions that involve only renormalizable couplings have been cross-checked with \texttt{RGBeta}~\cite{Thomsen:2021ncy} while the contributions with the $c_6$ coupling are found in agreement with~\cite{Cao:2021cdt,Cao:2023adc,Jenkins:2023bls}.

\section{Conclusions and outlook}

The discovery of physics beyond the SM is proving more challenging than initially anticipated. Given the precision increase associated with upcoming experimental searches and the absence of clear indications of the possible shape of NP, the use of novel and more precise EFT approaches becomes more important than ever. 

In this letter, we have presented the initial steps toward functional EFT matching and RG evolution at two-loop order, so far restricted to the case of scalar theories. We have explicitly demonstrated that the \emph{hard part} of the UV effective action is all that is needed for such calculations. To our knowledge, this is the first time that an explicit proof of this statement has been presented. While this result is applicable for both diagrammatic and functional approaches, it becomes particularly useful for the latter since it enables a power counting around which to perform an operator-product expansion of the functional results. Building on this, we have provided closed-form expressions for the evaluation of the two-loop effective action in the hard limit. In this way, and analogously to the one-loop functional result, our calculation of the two-loop EFT Lagrangian does not require the determination of the target EFT basis and essentially amounts to simple algebraic manipulations, making it particularly suitable for automation. We have also presented a toy-model example that illustrates the main rationale behind the application of our functional approach and, as a byproduct, we have verified recent literature results concerning the determination of two-loop RG equations in the geometric approach~\cite{Jenkins:2023bls}.

The extension of these methods to the more general case, including theories with fermions and/or gauge bosons, remains non-trivial and will be discussed in a forthcoming publication. In particular, we observe that the standard techniques to make the functional evaluations manifestly covariant~\cite{Gaillard:1985uh,Chan:1986jq} are no longer applicable at two-loop order, and new strategies are required. Likewise, the generalization of these results to higher-loop orders will also be explored in the future.

\section*{Acknowledgments}

We thank Julie Pag\`es and Supratim Das Bakshi for helpful discussions. The authors are grateful to the Mainz Institute for Theoretical Physics (MITP) of the Cluster of Excellence PRISMA$^+$ (Project ID 39083149) for its hospitality and support. The work of JFM is supported by the Spanish Ministry of Science and Innovation (MCIN) and the European Union NextGenerationEU/PRTR under grant IJC2020-043549-I, by the MCIN and State Research Agency (SRA) projects PID2019-106087GB-C22 and PID2022-139466NB-C21 (ERDF), and by the Junta de Andaluc\'ia projects P21\_00199 and FQM101. The work of AP has received funding from the Swiss National Science Foundation (SNSF) through the Eccellenza Professorial Fellowship ``Flavor Physics at the High Energy Frontier'' project number 186866. 
AET is funded by the Swiss National Science Foundation (SNSF) through the Ambizione grant ``Matching and Running: Improved Precision in the Hunt for New Physics,'' project number 209042.

\appendix
\renewcommand{\thesection}{\Alph{section}}
\renewcommand{\thesubsection}{\Alph{section}.\arabic{subsection}}

\section{Matching formula at two-loop order}
\label{app:proof}

In this appendix, we use the functional expansion of the effective action to demonstrate the explicit cancellation between soft-region loops in the UV and loops in the EFT and prove the master formula for EFT matching~\eqref{eq:gen_matching_formula} up to two-loop order. 

Our starting point is the off-shell matching condition~\eqref{eq:matching_condition}. On the UV side, we are dealing with the generating functional 
\begin{align} 
\Gamma^\phi_\UV[ \hat{\phi}] &\equiv \Gamma_\UV\big[ \widehat{\Phi}[\hat{\phi}],\, \hat{\phi}\big]\,,& 
\dfrac{\delta \Gamma_\sscript{UV}}{\delta \Phi_\alpha} \big[\widehat{\Phi}[\hat{\phi}],\, \hat{\phi}\big]=0\,,
\end{align}
which generates all 1LPI Green's functions. We need an explicit expression for $ \Gamma^\phi_\UV $ order-by-order in the loop expansion, and, so, we have to explicitly isolate each loop contribution in the heavy-field EOMs. To this end,  let $ \tilde{\eta}_I= \big(\overline{\Phi}_\alpha[\hat{\phi}],\, \hat{\phi}_i \big) $ solve the tree-level heavy field EOMs in the presence of $\hat{\phi}$, that is 
\begin{align} \label{eq:heavy_field_tree-level_eom}
\dfrac{\delta S_\UV^{(0)}}{\delta \Phi_\alpha}[\tilde{\eta}]=0\,.
\end{align}
Up to one-loop order, the quantum-EOM solution reads 
\begin{align}
\widehat{\Phi}_\alpha[\hat{\phi}] = \overline{\Phi}_\alpha[\hat{\phi}] - \hbar\,\widetilde{\mathcal{Q}}_{\alpha \beta}^{\eminus 1} \left( \widetilde{\mathcal{A}}^{(1)}_\beta + \dfrac{i}{2}\, \widetilde{\mathcal{C}}^{\,(0)}_{\beta IJ} \widetilde{\mathcal{Q}}_{JI}^{\eminus 1} \right) +\ord{\hbar^2}\,,
\end{align}
where we employ the same compact notation as in the main text, i.e. $\widetilde{\mathcal{Q}}_{IJ}=\mathcal{Q}_{IJ}[\tilde{\eta}]$ and likewise for other functions exclusively dependent of $\widetilde\eta_I$. Explicitly, the UV generating functional is then 
\begin{widetext}
\begin{align}\label{eq:1LPI_UV_eff_action}
\Gamma_\UV^{\phi}[\hat{\phi}] &= \widetilde{S}_\UV^{(0)} + \hbar\,\widetilde{S}_\UV^{(1)} + \dfrac{i\hbar}{2}  \big(\log \widetilde{\mathcal{Q}} \big)_{II} + \hbar^2\,\widetilde{S}_\UV^{(2)} + \dfrac{i\hbar^2}{2} \widetilde{\mathcal{Q}}_{IJ}^{\eminus1} \widetilde{\mathcal{B}}_{JI}^{(1)} + \dfrac{\hbar^2}{12} \widetilde{\mathcal{C}}_{IJK} \widetilde{\mathcal{Q}}_{IL}^{\eminus1} \widetilde{\mathcal{Q}}_{JM}^{\eminus1} \widetilde{\mathcal{Q}}_{KN}^{\eminus1} \widetilde{\mathcal{C}}_{LMN} \nonumber \\
&\quad - \dfrac{\hbar^2}{8} \widetilde{\mathcal{Q}}_{IJ}^{\eminus1}\,\widetilde{\mathcal{D}}_{IJKL} \widetilde{\mathcal{Q}}_{KL}^{\eminus1} - \dfrac{\hbar^2}{2} \bigg( \widetilde{\mathcal{A}}^{(1)}_\alpha+ \dfrac{i}{2}\widetilde{\mathcal{Q}}_{IJ}^{\eminus1} \,\widetilde{\mathcal{C}}^{(0)}_{IJ\alpha} \bigg)  \widetilde{\mathcal{Q}}_{\alpha\beta}^{\eminus1} \bigg( \widetilde{\mathcal{A}}^{(1)}_\beta+ \dfrac{i}{2} \,\widetilde{\mathcal{C}}^{(0)}_{\beta KL} \widetilde{\mathcal{Q}}_{KL}^{\eminus 1} \bigg)+\ord{\hbar^3}\,.
\end{align}
\end{widetext}
This expression is the same we would have gotten directly from~\eqref{eq:generating_functional} by setting $J_\Phi=0$ and performing the Legendre transform on the light fields only. We will also need the expression 
\begin{align} \label{eq:Phi0_dev}
\dfrac{\delta \overline{\Phi}_\alpha[\hat{\phi}]}{\delta \hat{\phi}_i } = -  \widetilde{\mathcal{Q}}_{i\beta} \widetilde{\mathcal{Q}}_{\beta \alpha }^{\eminus 1}\,,
\end{align}
for our proof, 
which is obtained by taking a functional derivative of the light classical fields in~\eqref{eq:heavy_field_tree-level_eom}. Finally, it is useful to define the upper-triangular transformation $ V_{IJ} $ that diagonalizes the inverse dressed propagator $\mathcal{Q}$. It takes the form  
\begin{align}
V_{IJ} &= 
\begin{pmatrix}
\delta_{\alpha \beta} & - \mathcal{Q}_{\alpha \gamma}^{\eminus 1} \mathcal{Q}_{\gamma j} \\
0 & \delta_{ij}   
\end{pmatrix}\,,
\end{align}
such that
\begin{align}
\mathds{Q}_{IJ} &\equiv V \transpose_{IK} \mathcal{Q}_{KL} V_{LJ} = 
\begin{pmatrix}
\mathcal{Q}_{\alpha \beta} & 0\\
0 & \mathcal{Q}_{ij}- \mathcal{Q}_{i \alpha} \mathcal{Q}^{\eminus 1}_{\alpha \beta} \mathcal{Q}_{\beta j}
\end{pmatrix}\,.
\end{align}
In the next sections, we proceed to establish the matching formula order by order in the loop expansion.

\subsection{Tree level}

It is important to discriminate between non-local objects such as $\mathcal{Q}^{\eminus 1}$ and their EFT operator-product expansions, as they have vastly different behaviors when performing the loop integrals. We denote the finite-order expansion of the heavy-field EOM solution by 
\begin{align}
\Phi^\mathrm{s} = \overline{\Phi}[\hat{\phi}] + \ord{1/ \Lambda^{N+1}} \,,
\end{align}   
which is sufficient to determine the $ N $-order matching. In what follows, the superscript `s' in any other quantities will be used to denote the dependence on $\Phi^\mathrm{s}[\hat{\phi}]$, e.g. $\mathcal{Q}^\mathrm{s} = \mathcal{Q}[ \eta^\mathrm{s}] = \mathcal{Q}\big[ \Phi^{\mathrm{s}}[\hat{\phi}], \hat{\phi} \big]$.

The tree-level matching criteria follows directly from the matching condition~\eqref{eq:matching_condition}, understood as a series in $ \Lambda $.  
At leading order, we simply find 
\begin{align}
S^{(0)}_\EFT[\hat{\phi}] = S_\UV^{(0)}[\eta^\mathrm{s}] =  S_\UV^{(0) \mathrm{s}} \,.
\end{align} 
This is a familiar result: the tree-level EFT is obtained by substituting in the UV theory the classic EOMs of all heavy fields expanded in the heavy masses.

\subsection{Expansion by regions} \label{app:exp_by_regions}

\begin{figure*}[t]
    \centering
    \includegraphics[width=\textwidth]{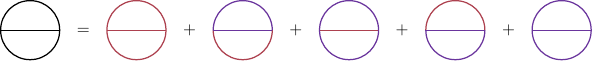}
    \caption{Region decomposition of the sunset graph, with red and violet lines denoting soft and hard momenta, respectively.}
    \label{fig:regions}
\end{figure*}

Loop integrals in dimensional regularization can, under certain conditions, be written as a sum over different momentum regions~\cite{Beneke:1997zp,Jantzen:2011nz}. This decomposition is known as expansion by regions, and in each region the propagators are series-expanded according to the magnitude of the loop momenta in that region. Working as we are, with a generic heavy scale set by $\Lambda$, we discriminate only between whether (Euclidean) loop momenta are soft ($\ll\Lambda$) or hard ($\gtrsim\Lambda$). Given a topology $G$, we can decompose it in regions as 
\begin{align} \label{eq:exp_by_regions_formula}
G= \sum_{\substack{\gamma \subseteq G\\ \mathrm{loops} } } R_\mathrm{hard}^{(\gamma) }(G)\,.
\end{align}
The sum runs over all (sub)loops $\gamma$ of $G$ and the operator $R_\mathrm{hard}^{(\gamma)}$ sets the corresponding loop momenta hard, while the remaining ones are taken to be soft. For instance, $\gamma$ runs over 5 subloops in the case of the sunset topology: no loops, 3 distinct one-loop graphs, and the full two-loop graph (see Figure~\ref{fig:regions}). 
We have verified that this region decomposition is valid up to two-loop order, following the criteria in~\cite{Jantzen:2011nz}.
To keep track of the regions in DeWitt notation, we introduce colored indices to distinguish soft and hard momentum modes:
\begin{align}
\soft{I}, \, \soft{\alpha},\, \soft{i}: \quad & p_\soft{I}, \, p_\soft{\alpha},\, p_\soft{i} \ll \Lambda\,, \nonumber\\
\hard{I}, \, \hard{\alpha},\, \hard{i}: \quad & p_\hard{I}, \, p_\hard{\alpha},\, p_\hard{i} \gtrsim \Lambda\,.
\end{align} 
All dressed propagators are understood as being expanded according to the color of their indices,  cf.~\eqref{eq:two_loop_topologies}.

In the EFT, there are no propagators with heavy masses, so all hard-momenta loops are scaleless and vanish in dimensional regularization. As a result, \emph{all indices in the EFT are soft type}. The series expansion in $\Phi^{\mathrm{s}}[\hat{\phi}]$ and in the soft regions are identical, so in general 
\begin{align} \label{eq:soft_region_func_devs}
\dfrac{\delta \Phi^\mathrm{s}_\alpha[\hat{\phi}]}{\delta \hat{\phi}_{\soft{i}}} = \left. \dfrac{\delta \overline{\Phi}^{(0)}_\alpha[\hat{\phi}]}{\delta \hat{\phi}_{\soft{i}}} \right|_{\overline{\Phi}^{(0)}= \Phi^\mathrm{s}}\,,
\end{align}
which, however, is not the case in the hard region (or for generic indices). Having defined this new notation, we turn to the one-loop case.

\subsection{One-loop level}

We use expansion by regions to decompose the one-loop UV effective action as
\begin{align}\label{eq:uv_1-loop_eff_action}
\Gamma_\UV^{\phi (1)}[\hat{\phi}] =	S_\UV^{(1) \mathrm{s}} + \dfrac{i}{2} \big(\!  \log \mathcal{Q}^\mathrm{s} \big)_{\!\hard{II}} + \dfrac{i}{2} \big(\!  \log \mathds{Q}^\mathrm{s} \big)_{\!\soft{ii}}\,. 
\end{align}
This follows from using the identity
\begin{align}
\mathrm{Tr} \log \mathcal{Q} = \log \mathrm{Det} \,\mathcal{Q}\,,
\end{align}
and block-diagonalizing the inverse dressed propagator with $V^\mathrm{s}$ satisfying $ \mathrm{Det}\,V = 1 $. The equality is then due to $ ( \log \mathds{Q}^\mathrm{s} )_{\soft{\alpha \alpha}} $ being a scaleless loop, as can be seen from $  \mathds{Q}^\mathrm{s}_{\alpha \beta } = \mathcal{Q}^\mathrm{s}_{\alpha \beta } $, indicating that heavy masses are present in all propagators.

Next, we determine the inverse dressed propagator in the EFT, as we need it for the EFT loop contributions. Using~\eqref{eq:soft_region_func_devs} and~\eqref{eq:Phi0_dev}, one can show that
\begin{align}\label{eq:eft_fluctuation_op}
\mathcal{Q}^{\EFT}_{\soft{ij}}[\hat{\phi}] = 
\dfrac{\delta^2 S_\UV^{(0)}[\eta^\mathrm{s}] }{\delta \hat{\phi}_\soft{i} \delta \hat{\phi}_\soft{j}} = \mathds{Q}^\mathrm{s}_\soft{ij}\,.
\end{align}
We can now determine the one-loop EFT action by inserting~\eqref{eq:eft_fluctuation_op} and~\eqref{eq:uv_1-loop_eff_action} into matching condition~\eqref{eq:matching_condition}: 
\begin{align} \label{eq:1-loop_matching_formula}
S_\EFT^{(1)}[\hat{\phi}] &= \Gamma_\UV^{\phi (1)}[\hat{\phi}] - \dfrac{i}{2} \big(\! \log \mathcal{Q}^{\EFT} \big)_{\soft{ii}} \nonumber\\
&= S_\UV^{(1) \mathrm{s}} + \dfrac{i}{2} \big(\!  \log \mathcal{Q}^\mathrm{s} \big)_{\!\hard{II}}\,.
\end{align}
This reproduces the well-known one-loop matching formula~\cite{Fuentes-Martin:2016uol,Zhang:2016pja}, and we can finally proceed to generalize it at the two-loop order.

\subsection{Two-loop level}

It will be useful to work in the basis where the inverse dressed propagator is block-diagonal also at two-loop order. To this end, we introduce the notation
\begin{align}
\mathds{A}^{(\ell)}_{I} &=\mathcal{A}^{(\ell)}_{J} V_{JI}\,,\nonumber\\
\mathds{B}^{(\ell)}_{IJ} &=\mathcal{B}^{(\ell)}_{KL} V_{KI} V_{LJ}\,,\nonumber\\
\mathds{C}^{(\ell)}_{IJK} &=\mathcal{C}^{(\ell)}_{LMN} V_{LI} V_{MJ} V_{NK}\,,\nonumber\\
\mathds{D}^{(\ell)}_{IJKL} &=\mathcal{D}^{(\ell)}_{MNOP} V_{MI} V_{NJ} V_{OK} V_{PL}\,.
\end{align}
Applying the expansion by regions~\eqref{eq:exp_by_regions_formula} to the 1LPI UV effective action~\eqref{eq:1LPI_UV_eff_action} yields
\begin{widetext} 
    \begin{align} \label{eq:Gamma2_UV_on-shell}
    \Gamma_\UV^\mathrm{\phi (2)}[\hat{\phi}] =&\phantom{+} S_\UV^{(2) \mathrm{s}} 
    + \dfrac{i}{2} \mathcal{Q}_{\hard{IJ}}^{\eminus1\, \mathrm{s}} \mathcal{B}^{(1)\mathrm{s}}_{\hard{JI}}
    +\dfrac{i}{2} \mathds{Q}_{\soft{IJ}}^{\eminus1\,\mathrm{s}} \mathds{B}_{\soft{JI}}^{(1) \mathrm{s}} -  \dfrac{1}{8} \mathcal{Q}^{\eminus1\, \mathrm{s}}_\hard{IJ} \mathcal{D}^{(0)\mathrm{s}}_\hard{IJKL} \mathcal{Q}^{\eminus1\, \mathrm{s}}_\hard{KL} -  \dfrac{1}{4} \mathds{Q}^{\eminus1\, \mathrm{s}}_\hard{IJ} \mathds{D}^{(0)\mathrm{s}}_{\hard{IJ}\soft{KL}} \mathds{Q}^{\eminus1\, \mathrm{s}}_\soft{KL} \nonumber\\
    & - \dfrac{1}{8} \mathds{Q}^{\eminus1\, \mathrm{s}}_\soft{IJ} \mathds{D}^{(0)\mathrm{s}}_{\soft{IJKL}} \mathds{Q}^{\eminus1\, \mathrm{s}}_\soft{KL}
    + \dfrac{1}{12} \mathcal{C}^{(0)\mathrm{s}}_\hard{IJK} \mathcal{Q}^{\eminus1\, \mathrm{s}}_\hard{IL} \mathcal{Q}^{\eminus1\, \mathrm{s}}_\hard{JM} \mathcal{Q}^{\eminus1\, \mathrm{s}}_\hard{KN} \mathcal{C}^{(0)\mathrm{s}}_\hard{LMN} + \dfrac{1}{4} \mathds{C}^{(0)\mathrm{s}}_{\soft{I} \hard{KL} } \mathds{Q}^{\eminus 1\, \mathrm{s}}_\soft{IJ} \mathds{Q}^\mathrm{\eminus 1 \,s}_{\hard{KM}} \mathds{Q}^\mathrm{\eminus 1 \,s}_{\hard{LN}} \mathds{C}^{(0)\mathrm{s}}_{\soft{J} \hard{MN}} \nonumber \\
    &
    +\dfrac{1}{12} \mathds{C}^{(0)\mathrm{s}}_\soft{IJK} \mathds{Q}^{\eminus 1\, \mathrm{s}}_\soft{IL} \mathds{Q}^{\eminus 1\, \mathrm{s}}_\soft{JM} \mathds{Q}^{\eminus 1\, \mathrm{s}}_\soft{KN} \mathds{C}^{(0)\mathrm{s}}_\soft{LMN} - \dfrac{1}{2} \bigg( \mathcal{A}^{(1) \mathrm{s}}_{\soft{\alpha}}+ \dfrac{i}{2} \mathcal{Q}_{\hard{IJ}}^{\eminus1\, \mathrm{s}} \mathcal{C}^{(0)\mathrm{s}}_{\hard{JI}\soft{\alpha}} \bigg)  \mathcal{Q}_{\soft{\alpha\beta}}^{\eminus1\, \mathrm{s}} \bigg( \mathcal{A}^{(1)\mathrm{s}}_{\soft{\beta}}+ \dfrac{i}{2} \mathcal{C}^{(0)\mathrm{s}}_{\soft{\beta} \hard{KL}} \mathcal{Q}_{\hard{LK}}^{\eminus 1\, \mathrm{s}} \bigg) \nonumber \\
    & - \dfrac{i}{2} \bigg(\mathds{A}^{(1) \mathrm{s}}_{\soft{\alpha}}+ \dfrac{i}{2} \mathds{Q}_{\hard{IJ}}^{\eminus1\, \mathrm{s}} \mathds{C}^{(0)\mathrm{s}}_{\hard{JI}\soft{\alpha}} \bigg)  \mathds{Q}_{\soft{\alpha\beta}}^{\eminus1\, \mathrm{s}} \mathds{C}^{(0)\mathrm{s}}_{\soft{\beta} \soft{KL}} \mathds{Q}_{\soft{LK}}^{\eminus 1\, \mathrm{s}} + \dfrac{1}{8} \mathds{Q}_{\soft{IJ}}^{\eminus1\, \mathrm{s}} \mathds{C}^{(0)\mathrm{s}}_{\soft{JI\alpha}} \mathds{Q}_{\soft{\alpha\beta}}^{\eminus1\, \mathrm{s}}  \mathds{C}^{(0)\mathrm{s}}_{\soft{\beta} \soft{KL}} \mathds{Q}_{\soft{LK}}^{\eminus 1\, \mathrm{s}}\,. 
    \end{align}
\end{widetext}
Consistently with the low-energy expansion, we have taken the soft limit in all tree-level propagators (present in the last three terms of the expression above). Building on the tree-level and one-loop EFT matching results, we can now construct the genuine loop contributions to the two-loop EFT effective action. EFT vertices are determined by the application of consecutive derivatives to the dressed propagator~\eqref{eq:eft_fluctuation_op} together with~\eqref{eq:Phi0_dev}:
\begin{align}
\mathcal{C}^{(0)\EFT}_\soft{ijk} [\hat{\phi}]&= \mathds{C}^{(0) \mathrm{s}}_\soft{ijk}\,, \nonumber\\*
\mathcal{D}^{(0)\EFT}_\soft{ijk\ell} [\hat{\phi}]&= \mathds{D}^{(0)\mathrm{s}}_\soft{ijk\ell} - \mathds{C}^{(0)\mathrm{s}}_\soft{ij\alpha} \mathds{Q}^{\eminus 1\, \mathrm{s}}_\soft{\alpha \beta} \mathds{C}^{(0)\mathrm{s}}_\soft{\beta k\ell} - \mathds{C}^{(0)\mathrm{s}}_\soft{ik\alpha} \mathds{Q}^{\eminus 1\, \mathrm{s}}_\soft{\alpha \beta} \mathds{C}^{(0)\mathrm{s}}_\soft{\beta j\ell} \nonumber \\*
& \quad - \mathds{C}^{(0)\mathrm{s}}_\soft{i\ell \alpha} \mathds{Q}^{\eminus 1\, \mathrm{s}}_\soft{\alpha \beta} \mathds{C}^{(0)\mathrm{s}}_\soft{\beta jk}\,.\\\nonumber
\end{align}
Hence, the EFT 4-point vertex is the 4-point function of the UV theory, with the heavy legs put on-shell with EOMs, supplemented by tree-level connected 4-point functions from a propagating heavy state in the UV. There are also contributions from the insertion of the one-loop EFT action in a one-loop topology. We find that
\begin{align}
\widehat{\mathcal{B}}^{(1) \EFT}_\soft{ij} &=  \mathds{B}^{(1)\mathrm{s}}_\soft{ij} - \left(\! \mathds{A}^{(1)\mathrm{s}}_{\soft{\alpha}}  + \dfrac{i}{2} \mathds{Q}^\mathrm{\eminus 1\, s}_{\hard{IJ}} \mathds{C}^{(0)\mathrm{s}}_{\hard{JI} \soft{\alpha}} \! \right)\!  \mathds{Q}^{\eminus 1\, s}_\soft{\alpha \beta } \mathds{C}^{(0)\mathrm{s}}_\soft{\beta ij}\nonumber\\
&\quad + \dfrac{i}{2} \mathds{Q}^\mathrm{\eminus 1\, s}_{\hard{IJ}} \mathds{D}^{(0)\mathrm{s}}_{\hard{JI} \soft{ij}} - \dfrac{i}{2} \mathds{C}^{(0)\mathrm{s}}_{\soft{i} \hard{KL} } \mathds{Q}^\mathrm{\eminus 1 \,s}_{\hard{KM}} \mathds{Q}^\mathrm{\eminus 1 \,s}_{\hard{LN}} \mathds{C}^{(0)\mathrm{s}}_{\hard{MN} \soft{j}}\,,
\end{align}
based on the matching result~\eqref{eq:1-loop_matching_formula}. Inserting these results into the expression for the two-loop effective action~\eqref{eq:eff_action_two-loop} and, by extending sums over light indices to run over all indices (recognizing that scaleless loop integrals vanish), we find the following result for the EFT effective action:
\begin{widetext}
\begin{align} \label{eq:Gamma2_EFT}
\Gamma^{(2)}_\EFT[\hat{\phi}] =& \phantom{+}\widehat{S}_\EFT^{(2)} + \dfrac{i}{2}  \mathds{B}^{(1)\mathrm{s}}_\soft{IJ} \mathds{Q}^{\eminus 1\, \mathrm{s}}_\soft{JI}
- \dfrac{i}{2} \left(\! \mathds{A}^{(1)\mathrm{s}}_{\soft{\alpha}}  + \dfrac{i}{2} \mathds{Q}^\mathrm{\eminus 1\, s}_{\hard{IJ}} \mathds{C}^{(0)\mathrm{s}}_{\hard{JI} \soft{\alpha}} \! \right)\!  \mathds{Q}^{\eminus 1\, s}_\soft{\alpha \beta } \mathds{C}^{(0)\mathrm{s}}_\soft{\beta IJ} \mathds{Q}^{\eminus 1\, \mathrm{s}}_\soft{JI}
- \dfrac{1}{4} \mathds{Q}^\mathrm{\eminus 1\, s}_{\hard{IJ}} \mathds{D}^{(0)\mathrm{s}}_{\hard{JI} \soft{KL}} \mathds{Q}^{\eminus 1\, \mathrm{s}}_\soft{LK} 
 \nonumber\\
&+ \dfrac{1}{4} \mathds{C}^{(0)\mathrm{s}}_{\soft{I} \hard{KL} } \mathds{Q}^{\eminus 1\, \mathrm{s}}_\soft{IJ} \mathds{Q}^\mathrm{\eminus 1 \,s}_{\hard{KM}} \mathds{Q}^\mathrm{\eminus 1 \,s}_{\hard{LN}} \mathds{C}^{(0)\mathrm{s}}_{\soft{J} \hard{MN}} -\dfrac{1}{8} \mathds{Q}^{\eminus 1\, \mathrm{s}}_\soft{IJ} \mathds{D}^{(0)\mathrm{s}}_\soft{IJKL} \mathds{Q}^{\eminus 1\, \mathrm{s}}_\soft{KL}
+ \dfrac{1}{8} \mathds{Q}^{\eminus 1\, \mathrm{s}}_\soft{IJ} \mathds{C}^{(0)\mathrm{s}}_\soft{IJ\alpha} \mathds{Q}^{\eminus 1\, \mathrm{s}}_\soft{\alpha \beta} \mathds{C}^{(0)\mathrm{s}}_\soft{\beta KL} \mathds{Q}^{\eminus 1\, \mathrm{s}}_\soft{KL} 
 \nonumber\\
&+\dfrac{1}{12} \mathds{C}^{(0)\mathrm{s}}_\soft{IJK} \mathds{Q}^{\eminus 1\, \mathrm{s}}_\soft{IL} \mathds{Q}^{\eminus 1\, \mathrm{s}}_\soft{JM} \mathds{Q}^{\eminus 1\, \mathrm{s}}_\soft{KN} \mathds{C}^{(0)\mathrm{s}}_\soft{LMN}\,.
\end{align}
We can now determine the master formula for two-loop matching directly from the matching condition~\eqref{eq:matching_condition} by canceling terms between~\eqref{eq:Gamma2_UV_on-shell} and~\eqref{eq:Gamma2_EFT}:
\begin{align} \label{eq:explicit_two-loop_matching_formula}
S_\EFT^{(2)}[\hat{\phi}] &= \Gamma_\UV^{\phi (2)}[\hat{\phi}] - \big(\Gamma_\EFT^{(2)}[\hat{\phi}] - \widehat{S}_\EFT^{(2)} \big)\nonumber\\
&= S_\UV^{(2)\mathrm{s}} + \dfrac{i}{2} \mathcal{Q}_{\hard{IJ}}^{\eminus1\, \mathrm{s}} \mathcal{B}_{\hard{JI}}^{(1)\, \mathrm{s}} -  \dfrac{1}{8} \mathcal{D}^{(0)\mathrm{s}}_\hard{IJKL} \mathcal{Q}^{\eminus1\, \mathrm{s}}_\hard{IJ} \mathcal{Q}^{\eminus1\, \mathrm{s}}_\hard{KL} + \dfrac{1}{12} \mathcal{C}^{(0) \mathrm{s}}_\hard{IJK} \mathcal{Q}^{\eminus1\, \mathrm{s}}_\hard{IL} \mathcal{Q}^{\eminus1\, \mathrm{s}}_\hard{JM} \mathcal{Q}^{\eminus1\, \mathrm{s}}_\hard{KN} \mathcal{C}^{(0) \mathrm{s}}_\hard{LMN}\nonumber\\
&\quad 	- \dfrac{1}{2} \bigg( \mathcal{A}^{(1) \mathrm{s}}_{\soft{\alpha}}+ \dfrac{i}{2} \mathcal{Q}_{\hard{IJ}}^{\eminus1\, \mathrm{s}} \mathcal{C}^{(0)\mathrm{s}}_{\hard{JI}\soft{\alpha}} \bigg)  \mathcal{Q}_{\soft{\alpha\beta}}^{\eminus1\, \mathrm{s}} \bigg( \mathcal{A}^{(1)\mathrm{s}}_{\soft{\beta}}+ \dfrac{i}{2} \mathcal{C}^{(0)\mathrm{s}}_{\soft{\beta} \hard{KL}} \mathcal{Q}_{\hard{LK}}^{\eminus 1\, \mathrm{s}} \bigg) \,.
\end{align}
\end{widetext}
We identify the master formula for two-loop matching~\eqref{eq:explicit_two-loop_matching_formula} as the two-loop case of the generic master formula~\eqref{eq:gen_matching_formula}. The first line of~\eqref{eq:explicit_two-loop_matching_formula} is the pure hard part of the two-loop 1PI topologies, whereas the second line comes from two insertion of the hard one-loop contribution to the heavy-field EOM.

\bibliographystyle{JHEP}
\bibliography{refs.bib}

\providecommand{\href}[2]{#2}\begingroup\raggedright\begin{thebibliography}{10}

\bibitem{Dawson:2022ewj}
S.~Dawson et~al., \emph{{LHC EFT WG Note: Precision matching of microscopic
  physics to the Standard Model Effective Field Theory (SMEFT)}},
  \href{https://arxiv.org/abs/2212.02905}{{\ttfamily 2212.02905}}.

\bibitem{Aebischer:2023irs}
J.~Aebischer et~al., \emph{{Computing Tools for Effective Field Theories}},
  \href{https://arxiv.org/abs/2307.08745}{{\ttfamily 2307.08745}}.

\bibitem{Henning:2014wua}
B.~Henning, X.~Lu and H.~Murayama, \emph{{How to use the Standard Model
  effective field theory}},
  \href{https://doi.org/10.1007/JHEP01(2016)023}{\emph{JHEP} {\bfseries 01}
  (2016) 023}, [\href{https://arxiv.org/abs/1412.1837}{{\ttfamily 1412.1837}}].

\bibitem{Fraser:1984zb}
C.~M. Fraser, \emph{{Calculation of Higher Derivative Terms in the One Loop
  Effective Lagrangian}}, \href{https://doi.org/10.1007/BF01550255}{\emph{Z.
  Phys. C} {\bfseries 28} (1985) 101}.

\bibitem{Aitchison:1984ys}
I.~J.~R. Aitchison and C.~M. Fraser, \emph{{Fermion Loop Contribution to
  Skyrmion Stability}},
  \href{https://doi.org/10.1016/0370-2693(84)90644-0}{\emph{Phys. Lett. B}
  {\bfseries 146} (1984) 63--66}.

\bibitem{Aitchison:1985pp}
I.~J.~R. Aitchison and C.~M. Fraser, \emph{{Derivative Expansions of Fermion
  Determinants: Anomaly Induced Vertices, Goldstone-Wilczek Currents and Skyrme
  Terms}}, \href{https://doi.org/10.1103/PhysRevD.31.2605}{\emph{Phys. Rev. D}
  {\bfseries 31} (1985) 2605}.

\bibitem{Aitchison:1985hu}
I.~J.~R. Aitchison and C.~M. Fraser, \emph{{Trouble With Boson Loops in
  Skyrmion Physics}},
  \href{https://doi.org/10.1103/PhysRevD.32.2190}{\emph{Phys. Rev. D}
  {\bfseries 32} (1985) 2190}.

\bibitem{Chan:1985ny}
L.~H. Chan, \emph{{Effective Action Expansion in Perturbation Theory}},
  \href{https://doi.org/10.1103/PhysRevLett.54.1222}{\emph{Phys. Rev. Lett.}
  {\bfseries 54} (1985) 1222--1225}.

\bibitem{Chan:1986jq}
L.-H. Chan, \emph{{Derivative Expansion for the One Loop Effective Actions With
  Internal Symmetry}},
  \href{https://doi.org/10.1103/PhysRevLett.57.1199}{\emph{Phys. Rev. Lett.}
  {\bfseries 57} (1986) 1199}.

\bibitem{Gaillard:1985uh}
M.~K. Gaillard, \emph{{The Effective One Loop Lagrangian With Derivative
  Couplings}}, \href{https://doi.org/10.1016/0550-3213(86)90264-6}{\emph{Nucl.
  Phys. B} {\bfseries 268} (1986) 669--692}.

\bibitem{Cheyette:1985ue}
O.~Cheyette, \emph{{Derivative Expansion of the Effective Action}},
  \href{https://doi.org/10.1103/PhysRevLett.55.2394}{\emph{Phys. Rev. Lett.}
  {\bfseries 55} (1985) 2394}.

\bibitem{Fuentes-Martin:2016uol}
J.~Fuentes-Martin, J.~Portoles and P.~Ruiz-Femenia, \emph{{Integrating out
  heavy particles with functional methods: a simplified framework}},
  \href{https://doi.org/10.1007/JHEP09(2016)156}{\emph{JHEP} {\bfseries 09}
  (2016) 156}, [\href{https://arxiv.org/abs/1607.02142}{{\ttfamily
  1607.02142}}].

\bibitem{Zhang:2016pja}
Z.~Zhang, \emph{{Covariant diagrams for one-loop matching}},
  \href{https://doi.org/10.1007/JHEP05(2017)152}{\emph{JHEP} {\bfseries 05}
  (2017) 152}, [\href{https://arxiv.org/abs/1610.00710}{{\ttfamily
  1610.00710}}].

\bibitem{Drozd:2015rsp}
A.~Drozd, J.~Ellis, J.~Quevillon and T.~You, \emph{{The Universal One-Loop
  Effective Action}},
  \href{https://doi.org/10.1007/JHEP03(2016)180}{\emph{JHEP} {\bfseries 03}
  (2016) 180}, [\href{https://arxiv.org/abs/1512.03003}{{\ttfamily
  1512.03003}}].

\bibitem{Ellis:2016enq}
S.~A.~R. Ellis, J.~Quevillon, T.~You and Z.~Zhang, \emph{{Mixed
  heavy\textendash{}light matching in the Universal One-Loop Effective
  Action}}, \href{https://doi.org/10.1016/j.physletb.2016.09.016}{\emph{Phys.
  Lett. B} {\bfseries 762} (2016) 166--176},
  [\href{https://arxiv.org/abs/1604.02445}{{\ttfamily 1604.02445}}].

\bibitem{Ellis:2017jns}
S.~A.~R. Ellis, J.~Quevillon, T.~You and Z.~Zhang, \emph{{Extending the
  Universal One-Loop Effective Action: Heavy-Light Coefficients}},
  \href{https://doi.org/10.1007/JHEP08(2017)054}{\emph{JHEP} {\bfseries 08}
  (2017) 054}, [\href{https://arxiv.org/abs/1706.07765}{{\ttfamily
  1706.07765}}].

\bibitem{Summ:2018oko}
B.~Summ and A.~Voigt, \emph{{Extending the Universal One-Loop Effective Action
  by Regularization Scheme Translating Operators}},
  \href{https://doi.org/10.1007/JHEP08(2018)026}{\emph{JHEP} {\bfseries 08}
  (2018) 026}, [\href{https://arxiv.org/abs/1806.05171}{{\ttfamily
  1806.05171}}].

\bibitem{DasBakshi:2018vni}
S.~Das~Bakshi, J.~Chakrabortty and S.~K. Patra, \emph{{CoDEx: Wilson
  coefficient calculator connecting SMEFT to UV theory}},
  \href{https://doi.org/10.1140/epjc/s10052-018-6444-2}{\emph{Eur. Phys. J. C}
  {\bfseries 79} (2019) 21},
  [\href{https://arxiv.org/abs/1808.04403}{{\ttfamily 1808.04403}}].

\bibitem{Kramer:2019fwz}
M.~Kr\"amer, B.~Summ and A.~Voigt, \emph{{Completing the scalar and fermionic
  Universal One-Loop Effective Action}},
  \href{https://doi.org/10.1007/JHEP01(2020)079}{\emph{JHEP} {\bfseries 01}
  (2020) 079}, [\href{https://arxiv.org/abs/1908.04798}{{\ttfamily
  1908.04798}}].

\bibitem{Ellis:2020ivx}
S.~A.~R. Ellis, J.~Quevillon, P.~N.~H. Vuong, T.~You and Z.~Zhang, \emph{{The
  Fermionic Universal One-Loop Effective Action}},
  \href{https://doi.org/10.1007/JHEP11(2020)078}{\emph{JHEP} {\bfseries 11}
  (2020) 078}, [\href{https://arxiv.org/abs/2006.16260}{{\ttfamily
  2006.16260}}].

\bibitem{Angelescu:2020yzf}
A.~Angelescu and P.~Huang, \emph{{Integrating Out New Fermions at One Loop}},
  \href{https://doi.org/10.1007/JHEP01(2021)049}{\emph{JHEP} {\bfseries 01}
  (2021) 049}, [\href{https://arxiv.org/abs/2006.16532}{{\ttfamily
  2006.16532}}].

\bibitem{Larue:2023uyv}
R.~Larue and J.~Quevillon, \emph{{The universal one-loop effective action with
  gravity}}, \href{https://doi.org/10.1007/JHEP11(2023)045}{\emph{JHEP}
  {\bfseries 11} (2023) 045},
  [\href{https://arxiv.org/abs/2303.10203}{{\ttfamily 2303.10203}}].

\bibitem{Banerjee:2023iiv}
U.~Banerjee, J.~Chakrabortty, S.~U. Rahaman and K.~Ramkumar, \emph{{One-loop
  Effective Action up to Dimension Eight: Integrating out Heavy Scalar(s)}},
  \href{https://arxiv.org/abs/2306.09103}{{\ttfamily 2306.09103}}.

\bibitem{Banerjee:2023xak}
U.~Banerjee, J.~Chakrabortty, S.~U. Rahaman and K.~Ramkumar, \emph{{One-loop
  Effective Action up to any Mass-dimension for Non-degenerate Scalars and
  Fermions including Light-Heavy Mixing}},
  \href{https://arxiv.org/abs/2311.12757}{{\ttfamily 2311.12757}}.

\bibitem{Chakrabortty:2023yke}
J.~Chakrabortty, S.~U. Rahaman and K.~Ramkumar, \emph{{One-loop Effective
  Action up to Dimension Eight: Integrating out Heavy Fermion(s)}},
  \href{https://arxiv.org/abs/2308.03849}{{\ttfamily 2308.03849}}.

\bibitem{Cohen:2020fcu}
T.~Cohen, X.~Lu and Z.~Zhang, \emph{{Functional Prescription for EFT
  Matching}}, \href{https://doi.org/10.1007/JHEP02(2021)228}{\emph{JHEP}
  {\bfseries 02} (2021) 228},
  [\href{https://arxiv.org/abs/2011.02484}{{\ttfamily 2011.02484}}].

\bibitem{Cohen:2020qvb}
T.~Cohen, X.~Lu and Z.~Zhang, \emph{{STrEAMlining EFT Matching}},
  \href{https://doi.org/10.21468/SciPostPhys.10.5.098}{\emph{SciPost Phys.}
  {\bfseries 10} (2021) 098},
  [\href{https://arxiv.org/abs/2012.07851}{{\ttfamily 2012.07851}}].

\bibitem{Fuentes-Martin:2020udw}
J.~Fuentes-Martin, M.~K\"onig, J.~Pag\`es, A.~E. Thomsen and F.~Wilsch,
  \emph{{SuperTracer: A Calculator of Functional Supertraces for One-Loop EFT
  Matching}}, \href{https://doi.org/10.1007/JHEP04(2021)281}{\emph{JHEP}
  {\bfseries 04} (2021) 281},
  [\href{https://arxiv.org/abs/2012.08506}{{\ttfamily 2012.08506}}].

\bibitem{vonGersdorff:2022kwj}
G.~von Gersdorff and K.~Santos, \emph{{New covariant Feynman rules for
  effective field theories}},
  \href{https://doi.org/10.1007/JHEP04(2023)025}{\emph{JHEP} {\bfseries 04}
  (2023) 025}, [\href{https://arxiv.org/abs/2212.07451}{{\ttfamily
  2212.07451}}].

\bibitem{vonGersdorff:2023lle}
G.~von Gersdorff, \emph{{Factorization of covariant Feynman graphs for the
  effective action}},  \href{https://arxiv.org/abs/2309.14939}{{\ttfamily
  2309.14939}}.

\bibitem{Carmona:2021xtq}
A.~Carmona, A.~Lazopoulos, P.~Olgoso and J.~Santiago, \emph{{Matchmakereft:
  automated tree-level and one-loop matching}},
  \href{https://doi.org/10.21468/SciPostPhys.12.6.198}{\emph{SciPost Phys.}
  {\bfseries 12} (2022) 198},
  [\href{https://arxiv.org/abs/2112.10787}{{\ttfamily 2112.10787}}].

\bibitem{Fuentes-Martin:2022jrf}
J.~Fuentes-Mart\'\i{}n, M.~K\"onig, J.~Pag\`es, A.~E. Thomsen and F.~Wilsch,
  \emph{{A proof of concept for matchete: an automated tool for matching
  effective theories}},
  \href{https://doi.org/10.1140/epjc/s10052-023-11726-1}{\emph{Eur. Phys. J. C}
  {\bfseries 83} (2023) 662},
  [\href{https://arxiv.org/abs/2212.04510}{{\ttfamily 2212.04510}}].

\bibitem{Jenkins:2013wua}
E.~E. Jenkins, A.~V. Manohar and M.~Trott, \emph{{Renormalization Group
  Evolution of the Standard Model Dimension Six Operators II: Yukawa
  Dependence}}, \href{https://doi.org/10.1007/JHEP01(2014)035}{\emph{JHEP}
  {\bfseries 01} (2014) 035},
  [\href{https://arxiv.org/abs/1310.4838}{{\ttfamily 1310.4838}}].

\bibitem{Jenkins:2013zja}
E.~E. Jenkins, A.~V. Manohar and M.~Trott, \emph{{Renormalization Group
  Evolution of the Standard Model Dimension Six Operators I: Formalism and
  lambda Dependence}},
  \href{https://doi.org/10.1007/JHEP10(2013)087}{\emph{JHEP} {\bfseries 10}
  (2013) 087}, [\href{https://arxiv.org/abs/1308.2627}{{\ttfamily 1308.2627}}].

\bibitem{Alonso:2013hga}
R.~Alonso, E.~E. Jenkins, A.~V. Manohar and M.~Trott, \emph{{Renormalization
  Group Evolution of the Standard Model Dimension Six Operators III: Gauge
  Coupling Dependence and Phenomenology}},
  \href{https://doi.org/10.1007/JHEP04(2014)159}{\emph{JHEP} {\bfseries 04}
  (2014) 159}, [\href{https://arxiv.org/abs/1312.2014}{{\ttfamily 1312.2014}}].

\bibitem{Bakshi:2021ofj}
S.~D. Bakshi, J.~Chakrabortty, C.~Englert, M.~Spannowsky and P.~Stylianou,
  \emph{{Landscaping CP-violating BSM scenarios}},
  \href{https://doi.org/10.1016/j.nuclphysb.2022.115676}{\emph{Nucl. Phys. B}
  {\bfseries 975} (2022) 115676},
  [\href{https://arxiv.org/abs/2103.15861}{{\ttfamily 2103.15861}}].

\bibitem{Ardu:2021koz}
M.~Ardu and S.~Davidson, \emph{{What is Leading Order for LFV in SMEFT?}},
  \href{https://doi.org/10.1007/JHEP08(2021)002}{\emph{JHEP} {\bfseries 08}
  (2021) 002}, [\href{https://arxiv.org/abs/2103.07212}{{\ttfamily
  2103.07212}}].

\bibitem{Allwicher:2023aql}
L.~Allwicher, G.~Isidori, J.~M. Lizana, N.~Selimovic and B.~A. Stefanek,
  \emph{{Third-family quark-lepton Unification and electroweak precision
  tests}}, \href{https://doi.org/10.1007/JHEP05(2023)179}{\emph{JHEP}
  {\bfseries 05} (2023) 179},
  [\href{https://arxiv.org/abs/2302.11584}{{\ttfamily 2302.11584}}].

\bibitem{Ciuchini:1993ks}
M.~Ciuchini, E.~Franco, G.~Martinelli, L.~Reina and L.~Silvestrini,
  \emph{{Scheme independence of the effective Hamiltonian for $b \to s\gamma$
  and $b \to s g$ decays}},
  \href{https://doi.org/10.1016/0370-2693(93)90668-8}{\emph{Phys. Lett. B}
  {\bfseries 316} (1993) 127--136},
  [\href{https://arxiv.org/abs/hep-ph/9307364}{{\ttfamily hep-ph/9307364}}].

\bibitem{Ciuchini:1993fk}
M.~Ciuchini, E.~Franco, L.~Reina and L.~Silvestrini, \emph{{Leading order QCD
  corrections to $b \to s\gamma$ and $b \to s g$ decays in three regularization
  schemes}}, \href{https://doi.org/10.1016/0550-3213(94)90223-2}{\emph{Nucl.
  Phys. B} {\bfseries 421} (1994) 41--64},
  [\href{https://arxiv.org/abs/hep-ph/9311357}{{\ttfamily hep-ph/9311357}}].

\bibitem{Jackiw:1974cv}
R.~Jackiw, \emph{{Functional evaluation of the effective potential}},
  \href{https://doi.org/10.1103/PhysRevD.9.1686}{\emph{Phys. Rev. D} {\bfseries
  9} (1974) 1686}.

\bibitem{Cornwall:1974vz}
J.~M. Cornwall, R.~Jackiw and E.~Tomboulis, \emph{{Effective Action for
  Composite Operators}},
  \href{https://doi.org/10.1103/PhysRevD.10.2428}{\emph{Phys. Rev. D}
  {\bfseries 10} (1974) 2428--2445}.

\bibitem{Gasser:1983yg}
J.~Gasser and H.~Leutwyler, \emph{{Chiral Perturbation Theory to One Loop}},
  \href{https://doi.org/10.1016/0003-4916(84)90242-2}{\emph{Annals Phys.}
  {\bfseries 158} (1984) 142}.

\bibitem{Ford:1992pn}
C.~Ford, I.~Jack and D.~R.~T. Jones, \emph{{The Standard model effective
  potential at two loops}},
  \href{https://doi.org/10.1016/0550-3213(92)90165-8}{\emph{Nucl. Phys. B}
  {\bfseries 387} (1992) 373--390},
  [\href{https://arxiv.org/abs/hep-ph/0111190}{{\ttfamily hep-ph/0111190}}].

\bibitem{Bijnens:1999hw}
J.~Bijnens, G.~Colangelo and G.~Ecker, \emph{{Renormalization of chiral
  perturbation theory to order $p^6$}},
  \href{https://doi.org/10.1006/aphy.1999.5982}{\emph{Annals Phys.} {\bfseries
  280} (2000) 100--139},
  [\href{https://arxiv.org/abs/hep-ph/9907333}{{\ttfamily hep-ph/9907333}}].

\bibitem{Beneke:1997zp}
M.~Beneke and V.~A. Smirnov, \emph{{Asymptotic expansion of Feynman integrals
  near threshold}},
  \href{https://doi.org/10.1016/S0550-3213(98)00138-2}{\emph{Nucl. Phys. B}
  {\bfseries 522} (1998) 321--344},
  [\href{https://arxiv.org/abs/hep-ph/9711391}{{\ttfamily hep-ph/9711391}}].

\bibitem{Jantzen:2011nz}
B.~Jantzen, \emph{{Foundation and generalization of the expansion by regions}},
  \href{https://doi.org/10.1007/JHEP12(2011)076}{\emph{JHEP} {\bfseries 12}
  (2011) 076}, [\href{https://arxiv.org/abs/1111.2589}{{\ttfamily 1111.2589}}].

\bibitem{Martin:2016bgz}
S.~P. Martin and D.~G. Robertson, \emph{{Evaluation of the general 3-loop
  vacuum Feynman integral}},
  \href{https://doi.org/10.1103/PhysRevD.95.016008}{\emph{Phys. Rev. D}
  {\bfseries 95} (2017) 016008},
  [\href{https://arxiv.org/abs/1610.07720}{{\ttfamily 1610.07720}}].

\bibitem{Dugan:1990df}
M.~J. Dugan and B.~Grinstein, \emph{{On the vanishing of evanescent
  operators}}, \href{https://doi.org/10.1016/0370-2693(91)90680-O}{\emph{Phys.
  Lett. B} {\bfseries 256} (1991) 239--244}.

\bibitem{Buras:1989xd}
A.~J. Buras and P.~H. Weisz, \emph{{QCD Nonleading Corrections to Weak Decays
  in Dimensional Regularization and 't Hooft-Veltman Schemes}},
  \href{https://doi.org/10.1016/0550-3213(90)90223-Z}{\emph{Nucl. Phys. B}
  {\bfseries 333} (1990) 66--99}.

\bibitem{Herrlich:1994kh}
S.~Herrlich and U.~Nierste, \emph{{Evanescent operators, scheme dependences and
  double insertions}},
  \href{https://doi.org/10.1016/0550-3213(95)00474-7}{\emph{Nucl. Phys. B}
  {\bfseries 455} (1995) 39--58},
  [\href{https://arxiv.org/abs/hep-ph/9412375}{{\ttfamily hep-ph/9412375}}].

\bibitem{Fuentes-Martin:2022vvu}
J.~Fuentes-Mart\'\i{}n, M.~K\"onig, J.~Pag\`es, A.~E. Thomsen and F.~Wilsch,
  \emph{{Evanescent operators in one-loop matching computations}},
  \href{https://doi.org/10.1007/JHEP02(2023)031}{\emph{JHEP} {\bfseries 02}
  (2023) 031}, [\href{https://arxiv.org/abs/2211.09144}{{\ttfamily
  2211.09144}}].

\bibitem{Aebischer:2022tvz}
J.~Aebischer, A.~J. Buras and J.~Kumar, \emph{{Simple rules for evanescent
  operators in one-loop basis transformations}},
  \href{https://doi.org/10.1103/PhysRevD.107.075007}{\emph{Phys. Rev. D}
  {\bfseries 107} (2023) 075007},
  [\href{https://arxiv.org/abs/2202.01225}{{\ttfamily 2202.01225}}].

\bibitem{Aebischer:2022aze}
J.~Aebischer and M.~Pesut, \emph{{One-loop Fierz transformations}},
  \href{https://doi.org/10.1007/JHEP10(2022)090}{\emph{JHEP} {\bfseries 10}
  (2022) 090}, [\href{https://arxiv.org/abs/2208.10513}{{\ttfamily
  2208.10513}}].

\bibitem{Aebischer:2022rxf}
J.~Aebischer, M.~Pesut and Z.~Polonsky, \emph{{Dipole operators in Fierz
  identities}},
  \href{https://doi.org/10.1016/j.physletb.2023.137968}{\emph{Phys. Lett. B}
  {\bfseries 842} (2023) 137968},
  [\href{https://arxiv.org/abs/2211.01379}{{\ttfamily 2211.01379}}].

\bibitem{Chetyrkin:1997fm}
K.~G. Chetyrkin, M.~Misiak and M.~Munz, \emph{{Beta functions and anomalous
  dimensions up to three loops}},
  \href{https://doi.org/10.1016/S0550-3213(98)00122-9}{\emph{Nucl. Phys. B}
  {\bfseries 518} (1998) 473--494},
  [\href{https://arxiv.org/abs/hep-ph/9711266}{{\ttfamily hep-ph/9711266}}].

\bibitem{Thomsen:2021ncy}
A.~E. Thomsen, \emph{{Introducing RGBeta: a Mathematica package for the
  evaluation of renormalization group $ \beta $-functions}},
  \href{https://doi.org/10.1140/epjc/s10052-021-09142-4}{\emph{Eur. Phys. J. C}
  {\bfseries 81} (2021) 408},
  [\href{https://arxiv.org/abs/2101.08265}{{\ttfamily 2101.08265}}].

\bibitem{Herren:2021yur}
F.~Herren and A.~E. Thomsen, \emph{{On ambiguities and divergences in
  perturbative renormalization group functions}},
  \href{https://doi.org/10.1007/JHEP06(2021)116}{\emph{JHEP} {\bfseries 06}
  (2021) 116}, [\href{https://arxiv.org/abs/2104.07037}{{\ttfamily
  2104.07037}}].

\bibitem{Cao:2021cdt}
W.~Cao, F.~Herzog, T.~Melia and J.~R. Nepveu, \emph{{Renormalization and
  non-renormalization of scalar EFTs at higher orders}},
  \href{https://doi.org/10.1007/JHEP09(2021)014}{\emph{JHEP} {\bfseries 09}
  (2021) 014}, [\href{https://arxiv.org/abs/2105.12742}{{\ttfamily
  2105.12742}}].

\bibitem{Cao:2023adc}
W.~Cao, F.~Herzog, T.~Melia and J.~Roosmale~Nepveu, \emph{{Non-linear
  non-renormalization theorems}},
  \href{https://doi.org/10.1007/JHEP08(2023)080}{\emph{JHEP} {\bfseries 08}
  (2023) 080}, [\href{https://arxiv.org/abs/2303.07391}{{\ttfamily
  2303.07391}}].

\bibitem{Jenkins:2023bls}
E.~E. Jenkins, A.~V. Manohar, L.~Naterop and J.~Pag\`es, \emph{{Two Loop
  Renormalization of Scalar Theories using a Geometric Approach}},
  \href{https://arxiv.org/abs/2310.19883}{{\ttfamily 2310.19883}}.

\end{thebibliography}\endgroup

\end{document}